\documentclass[a4paper,fleqn,usenatbib]{mnras}

\usepackage{newtxtext,newtxmath}

\usepackage[T1]{fontenc}
\usepackage{ae,aecompl}

\usepackage{graphicx}	
\usepackage{amsmath}	
\usepackage{amssymb}	
\usepackage{BibDef}

\usepackage{ragged2e}

\usepackage{booktabs}

\usepackage{multirow}

\usepackage{mathrsfs}

\newcommand{\msun}{\ensuremath{\, \mathrm M{\sun{}}}}

\title[Can SN kicks trigger EMRIs in the Galactic Centre?]%
{Can supernova kicks trigger EMRIs in the Galactic Centre?}

\author[Bortolas et al.]{
Elisa Bortolas$^{1,2,3}$\thanks{E-mail: elisa.bortolas@uzh.ch} \&
Michela Mapelli$^{3,2,4,5}$
\\
$^{1}$Center for Theoretical Astrophysics and Cosmology, Institute for Computational Science, University of Z\"urich, Winterthurerstrasse
 \\ \ 190, CH-8057 Z\"urich, Switzerland\\
$^{2}$INAF, Osservatorio Astronomico di Padova, Vicolo dell'Osservatorio 5, I-35122, Padova, Italy\\
$^{3}$Dipartimento di Fisica e Astronomia ``Galileo Galilei'', Universit\`a di Padova, Vicolo dell'Osservatorio 3, I-35122 Padova, Italy\\
$^{4}$Institut f\"ur Astro- und Teilchenphysik, Universit\"at Innsbruck, Technikerstrasse 25/8, A-6020, Innsbruck, Austria\\
$^{5}$INFN-Padova, Via Marzolo 8, I--35131 Padova, Italy\\
}
\date{Accepted XXX. Received YYY; in original from ZZZ}

\pubyear{2018}

\begin{document}
\label{firstpage}
\pagerange{\pageref{firstpage}--\pageref{lastpage}}
\maketitle

\begin{abstract}
One of the most promising gravitational wave (GW) sources detectable by the forthcoming LISA observatory are the so-called extreme-mass ratio inspirals (EMRIs), i.e. GW-driven inspirals of stellar-mass compact objects onto supermassive black holes (SMBHs). 
In this paper, we suggest that supernova (SN) kicks may trigger EMRIs in galactic nuclei by scattering 
newborn stellar black holes and neutron stars on extremely eccentric orbits; 
as a consequence,  the time-scale over which these compact objects  are expected to inspiral onto the central SMBH via GW emission may become shorter than the time-scale for other orbital perturbations to occur. By applying this argument to the Galactic Centre, we show that the S-cluster and the clockwise disc are optimal regions for the generation of such events: one SN out of $\sim10^4$ ($\sim10^5$) occurring in the S-cluster (clockwise disc) is expected to induce an EMRI. If we assume that the natal kicks  affecting stellar black holes are significantly slower than those experienced by neutron stars, we find that most SN-driven EMRIs involve neutron stars. We further estimate the time spanning from the SN to the final plunge onto the SMBH to be of the order of few Myr.   %
{Finally, we extrapolate the rate of SN-driven EMRIs per Milky Way to be up to 10$^{-8}$ yr$^{-1}$, thus we expect that LISA will detect up to a few tens of SN-driven EMRIs every year.}

\end{abstract}




\section{Introduction}
%

Extreme mass ratio inspirals (EMRIs) are  gravitational-wave (GW) driven decays of stellar-mass compact objects (COs) onto  supermassive black holes (SMBHs). 
In recent years, a growing body of  literature has been devoted to the generation of EMRIs (e.g. {%
\citealt{Levin2003, Miller2005, Hopman2005, Hopman2006, Amaro-Seoane2007, Gair2010, Merritt2011emris,Mapelli2012emri, 
Amaro-Seoane2013, Brem2014, Aharon2016,
Bar-Or2016, Babak2017, Chen2018%
}; see also the recent review by \citealt{Amaro-Seoane2018})}
as they could be detected in the near future by LISA, the space-borne GW interferometer selected by ESA for L3 \citep{Amaro-Seoane2017}. 
%
%
Tipically, EMRIs will shine in the milli-Hertz band of LISA completing up to $\sim 10^5$ cycles over the lifetime of the detector: such long-lived waveforms  will  exquisitely map the spacetime around SMBHs, and will provide unprecedented information on the SMBH masses, spins and  host environments {\citep{Gair2010, Gair2013, Barausse2014}}. 

In the standard picture, EMRIs are generated when a CO orbiting a SMBH is pushed on a sufficiently bound, low angular momentum orbit via two-body stellar scatterings \citep{Hils1995,Sigurdsson1997}; in fact, two-body relaxation is believed to constitute the main mechanism for the production of EMRIs. 
The associated event rate  has been inferred to  lie in the range  $10^{-9}-10^{-6}$ per galaxy per year; given that LISA will observe EMRIs up to $z\approx 2-3$, a few up to a few thousand EMRI signals are expected to { be detected by LISA every year \citep{Berry2016,Babak2017}. }

In this paper, we propose a novel, alternative mechanism for triggering EMRIs: through the supernova (SN) natal kick received by neutron stars (NSs) and stellar-mass black holes (BHs) at their birth. Circumstantial evidence suggests that NS kicks are very fast,  and considerably exceed the typical orbital velocity of stars in the field\footnote{We note that a slow-kick NS population may also exist \citep{Fryer1998, Pfahl2002, Arzoumanian2002, Beniamini2016,Verbunt2017, Mapelli2018, Giacobbo2018}.} { \citep{Hobbs2005, Bray2018, Igoshev2019}}. The kick distribution of BHs is much more debated, but theoretical and observational arguments suggest that, on average, BHs receive kicks that are  either below or comparable to the fast NS kicks \citep{Willems2005,Gualandris2005,Fragos2009,Repetto2012,Janka2013,Mandel2016,Repetto2017,Mirabel2017}.

When a SN kick occurs near a SMBH, the infant CO is likely  scattered on a completely different orbit, which might attain a very low angular momentum \citep{Bortolas2017}. Therefore, the newborn CO might gradually sink onto the SMBH via GW emission, reaching its ultimate coalescence before stellar orbital perturbations significantly deflect its trajectory.
In what follows, we will refer to these events as \textit{SN-EMRIs}. 

SN-EMRIs can solely occur in galactic nuclei  hosting  young ($\lesssim 50 $ Myr) and massive  ($\gtrsim9\msun$) stars whose life terminates in a SN explosion; furthermore,  SN-EMIRs can be detected by  LISA only if the inspiral occurs about a SMBH of $10^4-10^7 \msun$ at $z\lesssim 3$ \citep{Babak2017}. 
The closest nucleus to us, the Galactic Centre (GC), meets all the aforementioned requirements: it hosts a SMBH of $\approx4.3\times 10^6\msun$ \citep{Schodel2002,Ghez2003,Gillessen2017}, and several hundreds of young massive  stars (including Wolf-Rayet stars and O- and B-type stars) have been spotted within the  innermost 0.5 pc (\citealt{Paumard2006,Bartko2009,Do2013,Yelda2014}).
%
Furthermore, observations suggest that the recent ($\lesssim $ a few Myr, \citealt{Lu2013,Habibi2017}) star formation episode in the GC is associated to a top-heavy  mass function \citep{Lu2013}, implying that massive stars formed more effectively near the SMBH compared to the field. 
Owing to its vicinity, the GC is by far the best known galactic nucleus (see e.g. the review by \citealt{Mapelli2016}); thus we can model it in great detail. In particular, {observational evidence suggests that the Milky Way did not experience any recent  major merger \citep[e.g.][]{Wyse2001}, thus its nucleus can be assumed to be nearly dynamically relaxed \citep[e.g.][]{Baumgardt2017}}; furthermore, the Milky Way nuclear stellar cluster appears similar to other analogous regions observed in nearby galaxies \citep{Schodel2014} and can serve as a benchmark for the study of other nuclei.

Motivated by this, in the present paper we adopt a  Monte-Carlo approach to investigate the genesis of EMRIs triggered by SN kicks at the GC. Sec.~\ref{sec:theory} lays out the theoretical framework for SN-EMRIs production (\ref{sec:framework}), presents our modelling of the distribution functions adopted in the study (\ref{sec:distrib}) and describes our  numerical Monte-Carlo approach (\ref{sec:method}). Sec.~\ref{sec:results3} details the results of our investigation, which are then discussed in Sec.~\ref{sec:disc_concl}.

\section{Model and methods}\label{sec:theory}

\subsection{Theoretical framework}\label{sec:framework}

Let us suppose that a SMBH  of mass $M_\bullet$  sits at the origin of the coordinate system, and that a single massive star (of mass $m_i\ll M_\bullet$) orbits in a Keplerian fashion within its sphere of influence, with initial semi-major axis $a_i$ and eccentricity $e_i$. 
 When the star undergoes an SN, its  orbit gets significantly perturbed. For the sake of simplicity, the SN event is assumed to occur instantaneously: this allows to keep fixed the position  $\mathbf{r}$ (of modulus $r$) 
 at which the SN takes place; 
 the pre-SN Keplerian velocity at $\mathbf{r}$ is indicated with $\mathbf{v_i}$. 
 The SN kick can be mimicked by  adding a velocity vector $\mathbf{v_k}$, so that the newborn CO  velocity is $\mathbf{v} \equiv \mathbf{v_i} + \mathbf{v_k}$, and   lowering the pre-SN stellar mass $m_i$  to the post-SN, CO mass ($m_{\rm CO}$). 

After the kick,  the CO is still  bound to the SMBH if its energy per unit mass 
\begin{equation}
E=\frac 1 2 v^2-G\frac{M}{r}
\end{equation}
is negative; here $v^2= \mathbf{v\cdot v}$, $G$ is the gravitational constant and $M=M_\bullet+m_{\rm CO}(\approx M_\bullet)$.
If  $E>0$, the CO gets unbound from the SMBH and we assume it to be lost for the purpose of SN-EMRIs. If, instead, $E<0$, the new semi-major axis $a$ and eccentricity $e$ of the CO can be computed via
\begin{subequations}
\label{eq:neworbpar}
\begin{eqnarray}
a=-\frac{GM}{2E};\\
GM a(1-e^2)=|\mathbf{r}\times \mathbf{v}|^2.
\end{eqnarray}
\end{subequations}
{ Given the new orbital parameters, the characteristic timescale for energy loss  via gravitational radiation is \citep{Peters1964}:}
\begin{subequations}
\label{eq:tgw3}
\begin{eqnarray}
 t_{\rm GW}\approx \left. E \frac{dt}{dE} \right\rvert_{\rm GW} = \frac{5}{64}\frac{c^5a^4}{G^3M\,{}M_\bullet{}\,{}m_{\rm CO}}f(e);\\
 f(e)=(1-e^2)^{7/2}\left(1+\frac{73}{24}e^2+
\frac{37}{96}e^4\right)^{-1}\label{eq:fe},
\end{eqnarray}
\end{subequations}
{ where $\left.dE/dt \right\rvert_{\rm GW}$ indicates the rate of energy loss due to GWs and $c$ is  light speed}. Notably, $t_{\rm GW}$ strongly depends on the eccentricity. { The aforementioned time-scale} can be re-written as
\begin{equation}\begin{split}
t_{\rm GW}\approx 14 {\rm \  Gyr}\left(\frac{a}{3.3\times10^{-4}{\rm pc}}\right)^4 \left(\frac{M}{4.3\times10^{6}\msun}\right)^{-2}\\ \left(\frac{m_{\rm CO}}{25 \msun}\right)^{-1}
(1-e^2)^{7/2};
\end{split}
\end{equation}
in the assumption of $e\approx1$.   
Even if  $t_{\rm GW}$ has to be shorter than a Hubble time-scale (here assumed to be 14 Gyr) in order for the CO to be considered an EMRI candidate, this condition does not guarantee the inspiral to be successful, as repeated  two-body encounters with  background stars may irreparably perturb the CO orbit and suppress the GW-induced decay. 

\subsubsection{Two-body relaxation (NRR)} \label{sec:2br}


\begin{figure}
\centering
\includegraphics[ trim={3mm 0 3mm 0},clip,width=0.37\textwidth]{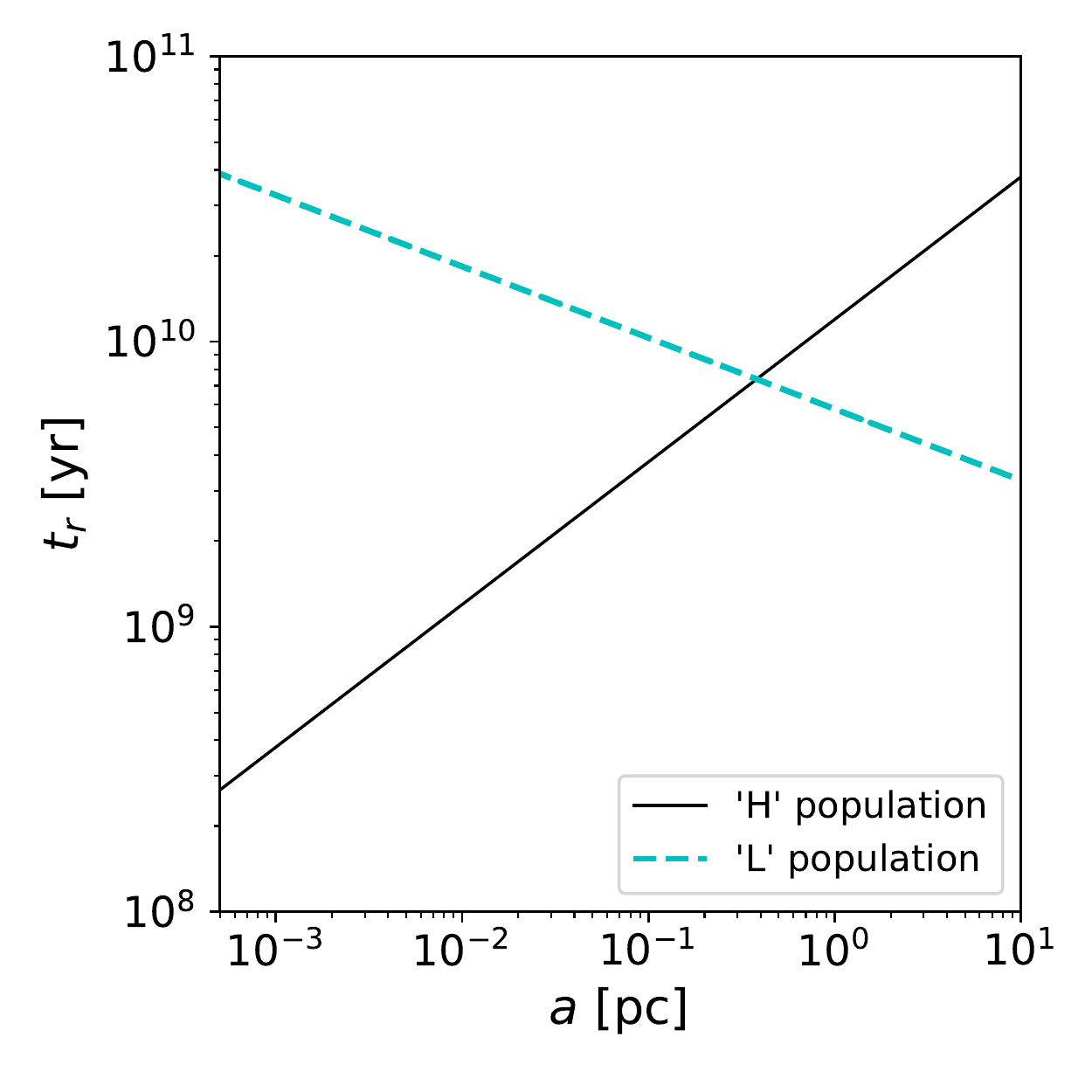}
\caption{ NRR timescale ($t_r$, computed via Eq.~\ref{eq:trel3}), as a function of the semimajor axis $a$ of stars orbiting the  $4.3\times 10^6\msun$ SMBH at the GC. The NRR timescale is evaluated assuming that relaxation effects are induced either by a steep cusp of relatively heavy stellar BHs (H population, solid black line) or by a shallower cusp populated by 1$\msun$ stars (L population, dashed cyan line); the quantities adopted for the computation of $t_r$ in the two cases are detailed in Tab~\ref{tab:stellar_pop}.}
\label{fig:trel3}
\end{figure}

\begin{figure}
\centering
\includegraphics[ trim={5mm 5mm 0mm 0},clip,width=0.48\textwidth]{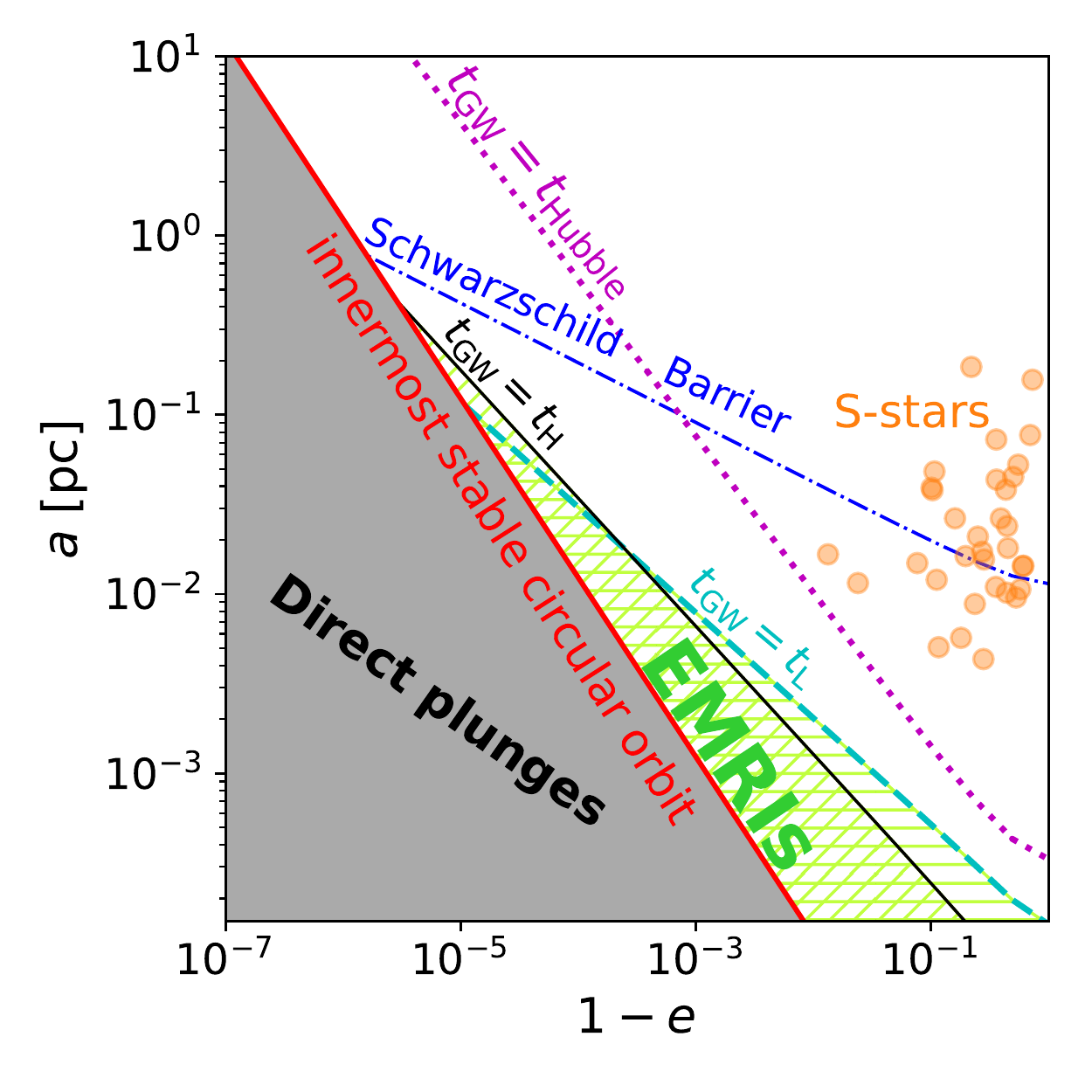} 
\caption{   $(1-e, a)$  space about the SMBH. COs scattered (either by NNR or via their natal kick) below the red solid line are classified as \textit{direct plunges}, as their periapsis distance falls inside the SMBH ISCO [i.e. $r_p\equiv a(1-e)<6r_g\equiv r_\bullet$].
The black solid (cyan dashed) line represents the outermost border of the region were proper EMRIs may occur according to Eq.~\ref{eq:aNRR}, assuming that the H (L) population dominates NRR at the GC. COs need to be scattered in the green hatched area to be classified as proper EMRIs. The plot further shows the phase space location of the S-stars (\citealt{Gillessen2017}, orange circles), excluding the eight S-stars classified as clockwise (CW) disc members; the violet dotted line delimits the region for which the GW inspiral timescale of a 10$\msun$-CO equals the Hubble time. Finally, the blue dash-dotted line displays the Schwarzschild barrier \citep[Eq.~\ref{eq:scwb}, ][]{Merritt2011emris} traced assuming  $C_{\rm SB}= 0.7$ and the properties of the H background population; we stress that {\it the Schwarzschild barrier does not affect the generation of SN-EMRIs}.
}\label{fig:fig123}
\end{figure}
%
\begin{table}
  \centering
  \caption{Properties of the two populations assumed  to dominate NRR at the GC}
  \label{tab:stellar_pop}
  \begin{center}
  
\begin{tabular}{lcccc}
\hline
Background pop. & $m_\star$ & $\gamma_{\rm NRR}$ & $N_{\rm NRR}(<1$ pc) & $t_r$(0.1 pc) \\ \hline
Light stars (`L') & 1 $\msun$ & 1.25 & $10^6$ & $1\times 10^{10}$ yr \\
Heavy stars (`H') & 10  $\msun$ & 2 & $10^4$ & $3.8\times 10^{9}$ yr \\ \hline
\end{tabular}

  \end{center}
\justifying
{\footnotesize  We report the quantities adopted in the computation of the NRR time in Eq.~\ref{eq:trel3}, exploring two possible stellar populations that dominate the GC relaxation effects: either a shallow cusp of standard, relatively light stars (L, first row), or a steeper cusp of heavier stellar BHs (H, second row). Each column indicates, for each population:  the typical stellar mass $m_\star$; the slope of the density profile, assumed to take the form  $n_{\rm NRR}(r)\propto r^{-\gamma_{_{\rm NRR}}}$; the number of objects $N_{\rm NRR}$ with semimajor axis shorter than 1 pc relative to the central SMBH;  the NRR timescale $t_r$ evaluated at $a=0.1$ pc. The adopted quantities are motivated in Sec.~\ref{sec:NRR_pop_distrib}.}
\end{table}

The effect of the two-body, non-resonant relaxation   { (NRR) induced by the accumulation of two-body encounters between stars } is two-folded: on the one hand, it may push stars on low angular momentum orbits and trigger the generation of  EMRIs  \citep{Hils1995,Sigurdsson1997,Amaro-Seoane2007}. On the other hand, and crucially in the SN-EMRI framework, NRR gradually perturbs the orbit of COs, thus EMRI candidates are likely to fail their inspiral if the duration of their GW decay exceeds the typical timescale over which NRR significantly deflects their trajectory. The time for gravitational two-body  encounters to change the angular momentum $j$ of a body of the order of itself is \citep{Hopman2005,Merritt2011emris}
\begin{equation}\label{eq:tL}
t_j=\left(\frac{j}{j_c}\right)^2t_r=(1-e^2)t_r\xrightarrow[e \to 1]{} 2(1-e)t_r  ,
\end{equation}
where $j_c$ is the angular momentum of an object with the same semi-major axis on a circular orbit, and $t_r$ indicates the NRR time-scale \citep{Spitzer1988}. Such latter quantity is best obtained by accounting for the fact that NRR acts more efficiently on very eccentric COs (i.e. those expected to undergo an EMRI, \citealt{Amaro-Seoane2007}) as implied in Eq.~\ref{eq:tL}: that is, low-angular-momentum bodies relax more efficiently%
\footnote{The diffusion in angular momentum resulting from NRR (addressed here) should not be confused with the so-called \textit{resonant relaxation}: the former results from uncorrelated two-body encounters with stars; the latter occurs as stars orbiting the SMBH in a Keplerian fashion swipe roughly  the same elliptical orbit over a long coherence period, maintaining a fixed relative orientation between each other; thus, in the latter case the gravitational interactions between stars are highly correlated, and produce an enhanced random walk of stars in angular momentum. An enlightening discussion on the mentioned mechanisms can be found in  sec.~2 of \citet{Bar-Or2016}.}%
. For this reason, we chose to compute the NRR time $t_r$ associated with angular momentum changes rather than the typically adopted one, evaluating energy changes. { To do so, we derived the orbit-averaged, Fokker-Plank diffusion coefficient describing NRR scattering in angular momentum and, following \citet{Hopman2005}, we define $t_r$ as the inverse of such coefficient. The obtained time-scale as a function of the semi-major axis $a$ of a generic star
within the SMBH influence sphere is
\begin{equation}\label{eq:trel3}
t_r=\frac{3\sqrt{2}\pi^2}{32 \,{}C_{\gamma_{\rm NRR}}} \left( \frac{G\,{}M_\bullet}{a}\right)^{3/2} \frac{a^{\gamma_{\rm NRR}}}{G^2 \,{}m_\star\,{} m'\,{} N_0\,{} \ln \Lambda};
\end{equation}
the full derivation of this equation can be found in Appendix~\ref{sec:appA}.} { Eq.~\ref{eq:trel3} has been obtained in the assumption that } stars (or stellar objects inducing angular momentum NRR) distribute isotropically and homogeneously 
about the SMBH, and their number density   scales as $n_{\rm NRR}(a)\propto a^{-\gamma_{\rm NRR}}$,  $\gamma_{\rm NRR}>0.5$; $N_0$ is the normalizing constant to the number $N_{\rm NRR}$ of stars within a given $a$, i.e. $N_{\rm NRR}(<a)=N_0a^{3-\gamma_{\rm NRR}}$. {  $C_{\gamma_{\rm NRR}}$ is a dimensionless constant of the order of unity in the cases of interest, whose derivation and analytic expression can be found in Appendix~\ref{sec:appA}. }
In Eq.~\ref{eq:trel3}, $m_\star$ represents the typical mass of stars that dominate NRR, while $m'$ is the mass of the  star undergoing  NRR. Finally, $ \ln \Lambda$ is the Coulomb logarithm, which can be approximated as $ \ln \Lambda=  \ln \left[ M_\bullet/\left(2m_\star\right)\right]$  within the SMBH sphere of influence (following \citealt{Merritt2011emris}).

The NRR time $t_r$ as a function of $a$ in  the GC  is displayed in  Fig.~\ref{fig:trel3}; we assumed NRR effects in the GC to be mainly induced either by a shallow cusp of relatively light stars (L population) or by a steep cusp of heavier stellar BHs (H population); the properties of the two groups are summarized in Tab.~\ref{tab:stellar_pop} and justified in Sec.~\ref{sec:NRR_pop_distrib}.

In order for a potential EMRI to complete its GW inspiral before NRR scatters it on a new trajectory,  its  GW decay time-scale ($t_{\rm GW}$, Eq.~\ref{eq:tgw3}) has to be shorter than $2(1-e)t_r$, with $t_r$ computed via Eq.~\eqref{eq:trel3}. Equating the two time-scales, it is possible to compute the limiting semi-major axis below which NRR is no longer efficient at quashing GW inspirals:
\begin{equation}\label{eq:aNRR}
a_{\rm NRR}^{-\gamma_{\rm NRR}+11/2} \approx \frac{6\sqrt{2}\pi^2}{5 C_{\gamma_{\rm NRR}}} \frac{1-e^2}{f(e)} \frac{(GM_\bullet)^{5/2}}{c^5} \frac{M_\bullet}{m_\star}\frac{1}{N_0\ln \Lambda}.
\end{equation}
The separatrix between the two regimes (Eq.~\ref{eq:aNRR}) is displayed in Fig.~\ref{fig:fig123} with a solid black (cyan dashed) line assuming the H (L) population to dominate NRR at the innermost GC parsec. If a CO sits at the left-hand side of such line (i.e., $a<a_{\rm NRR}$), NRR does not have enough time to perturb the CO orbit  prior to the final coalescence. 

\subsubsection{The `Schwarzschild barrier'}

It has been suggested that potential EMRIs induced by traditional NRR processes would hardly overcome the so-called Schwarzschild barrier, a boundary in phase space resulting from the coupling between resonant relaxation and relativistic precession \citep{Merritt2011emris}. The barrier can be parametrized via 
\begin{equation}\label{eq:scwb}
(1-e^2)_{\rm SB}= 1.9 \left( {C_{\rm SB}}/{0.7}\right)^2 \left( {r_g}/{a}\right)^2 \left( {M_\bullet}/{m_\star}\right)^2 {N_{\rm NRR}(<a)}^{-1},
\end{equation}
 where $C_{\rm SB}\approx 0.1-1$ \citep{Merritt2011emris,Brem2014} is an empirically determined constant, { $r_g$ is the gravitational radius of the SMBH,
\begin{equation}
r_g=\frac{GM_\bullet}{c^2},
\end{equation}
}and $N_{\rm NRR}(<a)$ is the number of bodies inside $a$ \citep{Merritt2011emris}.  The Schwarzschild barrier is shown as a blue dash-dotted line in Fig.~\ref{fig:fig123}. 
The presence of the barrier was originally believed to significantly suppress the rate of EMRIs induced by NRR \citep{Merritt2011emris}, but more recent studies suggest that the effect of the barrier is not as severe (\citealt{Amaro-Seoane2013,Brem2014}; see Sec. 7.6 of \citealt{Amaro-Seoane2018}). In any case, \textit{SN-EMRIs ignore the presence of the barrier}, as they are generated via a process other than NRR.

\subsubsection{Direct plunges}

In order for an EMRI to be successfully detected, the CO needs to swipe   multiple orbits while emitting detectable GWs. COs  directly plunging  into the SMBH innermost stable circular orbit (ISCO, $r_\bullet$), without undergoing a proper inspiral, are referred to as  {\it direct plunges}: they are expected to be swallowed by the SMBH without emitting significant GW radiation. Direct plunges are undetectable by LISA unless they they originate from the GC \citep{Hopman2005}. Even if they originate from the GC, the information contained in the signal would be very little.  
 If a SMBH is not rotating (Schwarzschild SMBH), then $r_\bullet=6r_g$. If the CO periapsis $r_p=a(1-e)$ falls inside $r_\bullet$, the event is classified as a direct plunge.

Traditionally, direct plunges have been distinguished from EMRIs by placing the direct-plunge line at $r_p=8r_g$, as highly eccentric orbits with $6r_g<r_p<8r_g$ emit significant GW only at periapsis, rather 
than emitting the continuously detectable, typical EMRI signal that best maps spacetime about the SMBH \citep[e.g.][]{Amaro-Seoane2017}; such highly eccentric sources have been named `extreme mass ratio bursts' \citep{Rubbo2006}. These events have been recently shown to be promising sources for LISA  if they originate from the GC or from a very nearby nucleus 
\citep{Berry2013,Berry2013b}. For this reason, here we set the delimiting line between plunges and EMRIs at $6r_g$.
The line separating plunges from other outcomes is shown as a red line in  Fig.~\ref{fig:fig123}. 

For the sake of completeness, we mention that when a SMBH is significantly rotating, $r_\bullet$ assumes different values depending on the dimensionless SMBH spin parameter $\mathcal{S}$, and on the orientation of the CO orbital angular momentum with respect to the SMBH spin. 
Notably, EMRI rates (versus plunge rates) were found to be boosted up to a factor of $\sim 30$  if the SMBH is significantly rotating \citep{Amaro-Seoane2013}. 
Here we will assume a Schwarzschild, non-rotating SMBH ($\mathcal{S}=0$, $r_\bullet=6r_g$), consistently with the probably slowly spinning SMBH at the GC ($\mathcal{S}\approx0.4$, \citealt{Kato2010}), as no substantial enhancement in the EMRI  rates is expected if the SMBH spin is $\mathcal{S}\lesssim0.7$ \citep{Amaro-Seoane2013}. 
\\

To sum up, a proper SN-EMRI can occur if the SN kick ejects the newborn CO to an orbit whose  semi-major axis is smaller than the threshold value $a_{\rm NRR}$ (Eq.~\ref{eq:aNRR}) at which NRR effects would deflect its orbit prior to  coalescence, and whose pericentre is greater than  
%
$r_p=6r_g$. We will refer to this area as the `EMRI' region of phase space (the green hatched area in Fig.~\ref{fig:fig123}).

\subsubsection{Further rejection criteria}\label{sec:further_rejection}
Depending on the progenitor star properties (i.e. its $m_i, a_i, e_i$) the SN-EMRI could  be aborted prior to the SN event. Here we list the abortion scenarios we accounted for in this study.

\begin{enumerate}
\item If the progenitor star initially orbits very near the SMBH, its GW inspiral timescale (Eq.~\ref{eq:tgw3}, using $m_i$ instead of $m_{\rm CO}$) can be shorter than the time spanning from the star birth to its SN explosion. If this happens, the object is lost into the SMBH prior to experiencing the SN.

\item Analogously, if the progenitor star pericentre $r_{p,i}=a_i(1-e_i)$ is smaller than its tidal disruption radius, i.e.
$r_{p,i}<({M_\bullet}/{m_i})^{1/3}R_{\rm strip}$, the star undergoes tidal disruption and is  lost for the purpose of SN-EMRIs.  { Even if stellar radii can become be much larger than the solar radius $R_\odot$ (e.g. in the case of red giant stars), a star can be considered to be successfully disrupted only if the radius of its core (rather than its envelope) undergoes tidal disruption. For this, we set $R_{\rm strip}$ to be equal to $3R_\odot$, i.e. roughly the maximum value that a helium-core radius can reach  \citep{Chen2015}.}
\end{enumerate}
Referring to point (ii) above, the progenitor star could survive tidal disruption, while being stripped solely of its outer envelope. Such naked stellar core can be identified  by checking whether $r_{p,i}<({M_\bullet}/{m_i})^{1/3}R_{\rm max}$, where $R_{\rm max}$ represents the maximum radius attained by the star during  its stellar evolutionary phases. We consider naked stellar cores to be valid SN-EMRI progenitors and we treat them as standard progenitor stars, although the missing envelope might affect their natal kick.

\subsection{Model} \label{sec:distrib}
In order to explore the generation of SN-EMRIs in the GC, one has to assume how stellar masses, orbits and SN velocity kicks of the young stellar population distribute around the SMBH. For this, in what follows we detail the distributions we adopted for our study.

\subsubsection{Stellar mass function and stellar evolution}\label{sec:mass_distrib}
The zero age main sequence (ZAMS) mass of the CO progenitor star is distributed between $9\msun$ and $100 \msun$ according to a top-heavy mass function in the form $dN/dm\propto m^{-\alpha}$, with $\alpha=1.7$, in agreement with the recent observations of young stars in the GC \citep{Lu2013}. We also verified that our results do not change significantly if we assume a \citet{Kroupa2001} mass function.

For any given ZAMS mass,  the PARSEC stellar evolutionary tracks at solar metallicity \citep{Bressan2012, Chen2015} allowed us to obtain the progenitor mass $m_i$ just prior to the SN; the final CO mass is then computed assuming a delayed SN explosion model \citep{Fryer2012}, following \citet{Spera2015}\footnote{The zero-age main sequence mass versus $m_i$ and $m_{\rm CO}$ is shown in fig. ~1 of \citet{Bortolas2017}.}. If $m_{\rm CO}<3 \msun  \ \ (\geq3 \msun)$, the CO is classified as a NS (a BH). The PARSEC tracks are also used to compute the time elapsed from the ZAMS to the SN event, and to estimate the maximum radius $R_{\rm max}$ attained by each star during its entire life.

\subsubsection{Initial orbit of the progenitor star}\label{sec:ae_distrib}

The distribution of the semi-major axis $a_i$ and eccentricity $e_i$ of stars undergoing a core-collapse SN are chosen assuming that the young stars in the GC  belong to one of the structures described below.

\begin{table*}
  \centering
  \caption{Initial conditions of the explored CO progenitor populations assumed in the GC}
  \label{tab:ic3}
  \begin{center}
  
\begin{tabular}{@{}lcccccl@{}}\hline
Population name & Label & $a_{i,\rm min}$ {[}pc{]} & $a_{i,\rm max}$ {[}pc{]} & $\gamma_i$ & ecc. distr. &  BH kicks \\ \hline
CW disc & CWD & 0.040 & 0.13 & 1.93 & GE & fast/slow \\
S-cluster & SCL & 0.001 & 0.04 & 1.10 & TE & fast/slow \\
concentrated profile, GE & CP-G & 0.001 & 1.00 & 2.00 & GE & fast/slow \\
concentrated profile, TE & CP-T & 0.001 & 1.00 & 2.00 & TE & fast/slow \\
shallow profile, GE & SP-G & 0.001 & 1.00 & 1.25 & GE & fast/slow \\
shallow profile, TE & SP-T & 0.001 & 1.00 & 1.25 & TE & fast/slow \\ 
 \hline
\end{tabular}

  \end{center}
  \justifying
{\footnotesize %
Each row lists the parameters adopted for describing the different young populations in the GC that undergo SN kicks. We modelled the observed clockwise (CW) disc of stars and  S-star cluster, and we further explored a  generic cuspy distribution for young stars at the GC. The first column reports the name of the population of progenitor stars; the second column shows the label we used to refer to each population throughout the paper; the  third and fourth columns show the minimum and maximum value of the semi-major axis; the fifth column lists the slope of the density profile of the young stars,  $n_i(a_i)\propto a_i^{-\gamma_i}$; the sixth column displays the eccentricity distribution adopted for each population, either a thermal one [TE, $n_e(e_i)\propto e_i$] or a Gaussian distribution (GE) centred at $<e_i>=0.3$ and with dispersion of 0.1; finally, the last column indicates that the SN kicks imparted to stellar BHs have been both not normalized (`fast') or normalized (`slow') to the final CO mass (more details are in the text). }
\end{table*}

\begin{figure}
\centering
\includegraphics[trim={3mm 0 3mm 0}, clip, width=0.38\textwidth]{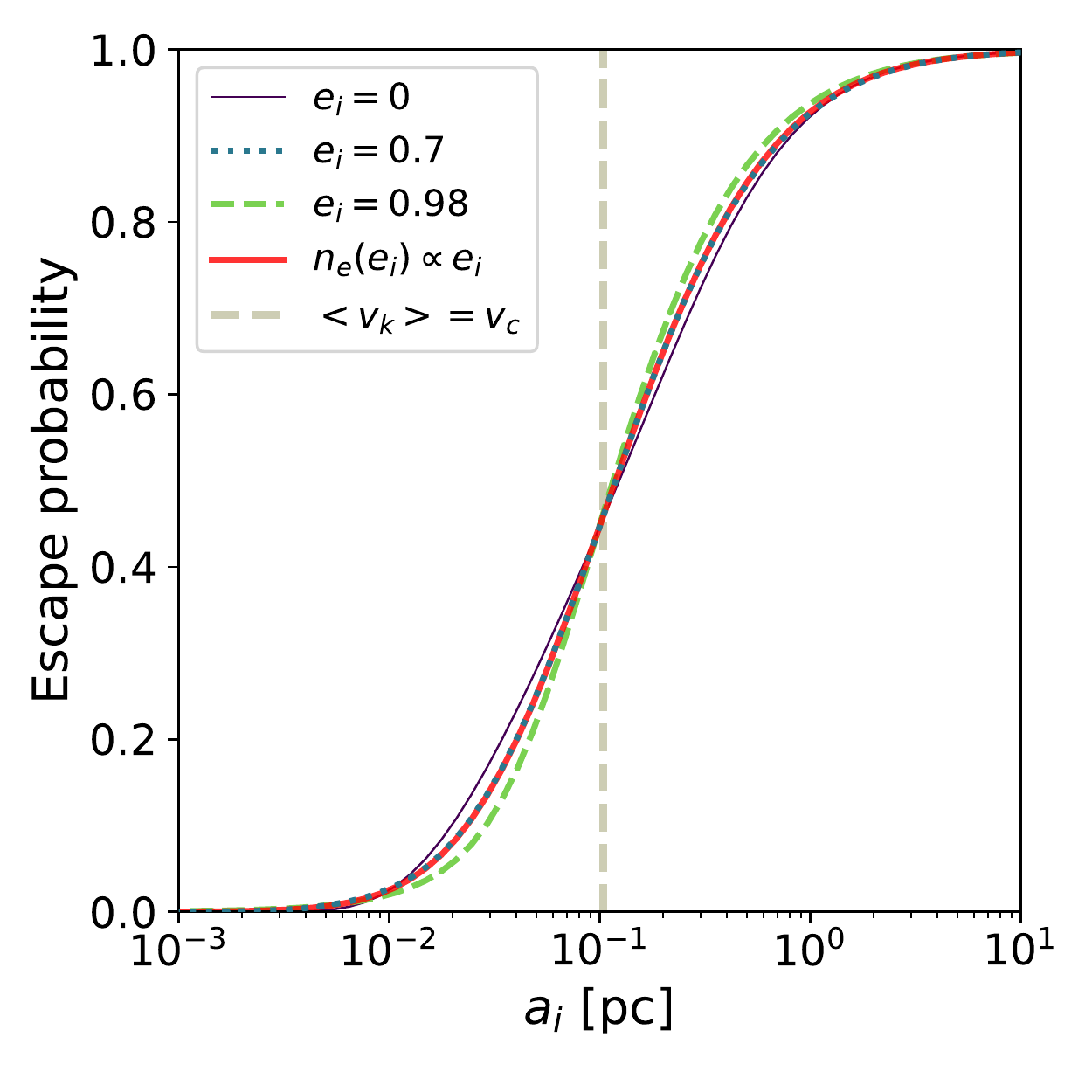} 
    \caption{The plot displays the probability of a CO to get unbound from the SMBH as a result of its natal kick, as a function of the semi-major axis of the progenitor star $a_i$. We only account for the SMBH potential and we neglect the presence of the nuclear stellar cluster about the SMBH. We model each $a_i$ as a Dirac delta distribution; different lines show different assumptions for the initial eccentricity at each $a_i$; $e_i$ is set equal to 0, 0.7, 0.98 (the violet solid, blue dotted and green dashed line respectively) or assumed to distribute with a TE (red solid line). We assume the `fast' prescription for SN kicks \citep{Hobbs2005}.  The probability for each object to get unbound by the SN kick exceeds  50\% at $a_i\approx 0.1034$ pc (vertical dashed grey line), at which the circular Keplerian velocity equals the mean value of the SN kick distribution ($\approx423$ km s$^{-1}$).}
    \label{fig:prob_e}
\end{figure}

\begin{enumerate}

\item {\it The CW disc.} The clockwise (CW) disc is an eccentric discy structure of a few $\times 10^4 \msun$  populated by young and massive (B-type, Wolf-Rayet) stars occupying the GC innermost fraction of parsec \citep{Lu2013}. According to recent observations, we modelled the CW disc as a region extended between $0.04$ and $0.13$ pc \citep{Do2013}.  The eccentricity distribution in the CW disc was set as a Gaussian centred in $<e_i>=0.3$ and with dispersion equal to 0.1 (\citealt{Yelda2014}; we will refer to the aforementioned Gaussian eccentricity distribution as GE throughout the paper), while the semi-major axis distribution was set as a power law in the form $f(a_i)\propto a_i^{0.07}$. This latter choice ensures that the surface density profile in the disc follows the observed $\Sigma(r)\propto r^{-0.93}$ \citep{Do2013}.

\item {\it The S-star cluster.} This is a cluster of at least $\sim 50$ stars (mostly of B stellar type) isotropically distributed in the innermost arcsecond from the GC SMBH \citep{Morris1993,Gillessen2013}. We modelled the S-stars semi-major axis distribution so that their density profile follows the observed $n_i(a_i)\propto a_i^{-1.1}$  in the range [0.001, 0.04] pc  \citep{Gillessen2009,Gillessen2017}. The eccentricity distribution of S-stars is observed to be thermal \citep{Gillessen2017}. Accordingly, we represented  S-star eccentricities as  $n_e(e_i)\propto e_i$. In what follows, we will refer to the thermal eccentricity distribution as TE.

\item {\it A generic stellar profile.} Additionally, we  explored the possibility that CO progenitors in the GC follow a density profile in the form  $n_i(a_i) \propto a_i^{-\gamma_i}$  between $0.001 - 1$ pc. The smallest allowed $a_i$ is comparable to the minimum observed semi-major axis among the S-stars (S55, with $a\approx0.0043$ pc,  \citealt{Gillessen2017}). The outer cluster limit is set to 1 pc, even if the influence sphere of the SMBH in the GC reaches $\approx 3$ pc \citep{Schodel2017,Gallego-Cano2017}: first of all, young stars in the GC are centrally concentrated, and 90\% of them seem to be found in the innermost 0.5 pc \citep{Feldmeier-Krause2015}; secondly, stars orbiting the SMBH at larger distances are very likely to get unbound from the SMBH as a result of their natal kick, as shown in Fig~\ref{fig:prob_e}. 

Two slopes of the density profile  have been explored for the progenitor stars\footnote{We  stress that the initial distribution of  young stars undergoing a SN is assumed to be unrelated to the properties of the underlying, older stellar population inducing NRR.}: $\gamma_i=1.25$ and $\gamma_i=2$, referred to as the concentrated and shallow density profile, respectively. We explored two different eccentricity distributions for each: a TE and a GE distribution.
\end{enumerate}

Tab.~\ref{tab:ic3} summarizes the parameters of the studied progenitor populations and reports the associated labels we used throughout the paper. Given that the assumption of a Gaussian rather than a thermal distribution for the eccentricities in our initial conditions significantly affects our results, we will refer to the progenitor structures with an underlying thermal distribution as the \textit{TE cases}, while the progenitor structures assuming the  Gaussian  distribution are referred to as the \textit{GE cases}.

In some cases, we further adopted  a Dirac delta function for distributing $a_i$ or $e_i$ or both, in order to disentangle the effect of the progenitor orbital parameters on the obtained rates of SN-EMRIs.

\subsubsection{Supernova kicks}\label{sec:kick_distrib}
The SN kicks are assumed to distribute isotropically in space; the magnitude of the SN kick imparted to the progenitor star, $v_k$, is drawn from a Maxwellian distribution with one-dimensional root-mean square velocity $\sigma=265$ km s$^{-1}$ in the case of NSs, according to  \citet{Hobbs2005}. Owing to the large uncertainties on BH natal kicks, we adopt two different approaches for BHs: (i) `fast' BH kicks: BH kicks are drawn from the same distribution as NSs; (ii) `slow' BH kicks: BH kicks are drawn from the same distribution as NSs, but are normalized to the final mass of the BH, assuming linear momentum conservation, as described in \citet{Bortolas2017}. 

\subsubsection{Stellar population dominating NRR effects}\label{sec:NRR_pop_distrib}
As anticipated above, we selected two different \textit{old stellar populations} that may dominate the NRR effects (i.e., the computation of $t_r$) in our model of the GC: (i) light stars with average mass  $m_\star=1\msun$, slope of the density profile $\gamma_{\rm NRR}=1.25$ and number within the inner parsec $N_{\rm NRR}(<1\ {\rm pc})=10^6$ (`L' population, \citealt{Schodel2017,Gallego-Cano2017,Baumgardt2017, Habibi2019}); (ii) stellar-mass BHs with average mass $m_\star=10\msun$, with a steeper density profile ($\gamma_{\rm NRR}=2$) and for which  $N_{\rm NRR}(<1\ {\rm pc})=10^4$ (`H' population, \citealt{Baumgardt2017,Hailey2018}). 

Our choice to explore two different populations is motivated by the fact that the density distribution within the inner parsec of the GC is hardly inferred via infra-red observations, and probably strongly affected by the presence of a massive, invisible cusp of dark objects (see e.g. the recent observational results by \citealt{Hailey2018}, and the theoretical work by \citealt{Baumgardt2017}); such dark cusp is expected to result from mass segregation \citep{Baumgardt2017}, possibly aided by the fact that BHs might receive slow natal kicks \citep[e.g.][]{Bortolas2017}. For this reason, we consider the `H' population our fiducial case.

\subsection{Methodology}\label{sec:method}
Here we sum-up the steps we follow in our Monte Carlo approach to study SN-EMRIs.
\begin{enumerate}
\item We place the CO progenitor star, whose mass $m_i$ is distributed as described in Sec.~\ref{sec:mass_distrib}, in a Keplerian bound orbit about a $M_\bullet = 4.3\times 10^6 \msun$ SMBH. The star initial orbital parameters $a_i$ and $e_i$ are chosen according to the distributions presented in Sec.~\ref{sec:ae_distrib} and in Tab.~\ref{tab:ic3}. We reject  stars whose orbital properties lead to a disruption event by the SMBH (as described in Sec.~\ref{sec:further_rejection}).
The exact point along the orbit at which the SN explosion occurs is obtained by drawing an orbital mean anomaly (which is uniformly distributed in time) from a uniform distribution in $[0,2\,{}\pi]$.
\item We add the SN velocity kick to the Keplerian velocity vector of the star, following the distribution described in Sec.~\ref{sec:kick_distrib}. We also turn the progenitor mass into the final CO mass according to the prescriptions in Sec.~\ref{sec:mass_distrib}.
\item We compute the new energy of the CO, and we check if it is still bound to the SMBH%
\footnote{It is worth stressing that objects that get unbound from the SMBH may remain bound to the galactic potential and get back to the SMBH at later times, even if we do not account for this possibility in the present study. However, we expect these objects will no longer be able to undergo an EMRI; rather, they might experience a direct plunge.}%
. If so, we compute its new orbital parameters ($a,e$) via Eq.~\ref{eq:neworbpar}, and we check whether the CO can be classified as an EMRI [if $6r_g/(1-e)<a< a_{\rm NRR}$] or a direct plunge  [if $a(1-e)<6r_g$].
\end{enumerate}

For each  of the  explored initial conditions (Tab.~\ref{tab:ic3}) we performed a large number of Monte Carlo iterations to filter out statistical fluctuations.

\subsection{Model limitations}
Our investigation suffers from a number of limitations. First of all, we neglect the possibility that stars exploding as SNe may be members of a binary system; this aspect is important, especially considering that the binary fraction in the CW disc probably lies in the range $30-85\%$ 
(\citealt{Pfuhl2014,Gautam2017};  { see also  \citealt{Hopman2009} for comprehensive study about the survival of binaries near a SMBH}). As a matter of fact, \citet{Lu2018} explore the dynamics of SN kicks in triple systems,  the third body possibly being an SMBH. They touch upon the possibility of producing LISA sources via SN explosions occurring in stellar binaries, but a more focussed study would be needed to address this aspect in detail. 

A second source of uncertainty is the fact that the SN kick velocity distribution remains largely debated, especially for BHs (e.g. \citealt{Beniamini2016} and references therein). Reassuringly, we find that the two different BH kick prescriptions explored in this study typically lead to the same order of magnitude of SN-EMRIs, although the SN-EMRI mass spectrum is significantly affected by the kick prescriptions.

{ In addition},  our model cannot account for  secular processes. Among them,  Kozai-Lidov resonances  \citep{Kozai1962,Lidov1962}  induced by the CW disc may produce oscillations in the eccentricity and inclination of a candidate SN-EMRI, and potentially affect its inspiral. However, the Kozai-Lidov timescale becomes shorter than the typical time over which a SN-EMRI is expected to complete its inspiral (1 Myr) only at $a\gtrsim 1$ pc (see e.g. fig.~23 in \citealt{Mapelli2016}), where only direct plunges can occur. Furthermore, resonant relaxation could  affect the inspiral of a CO, but this effect is expected to be more relevant than the addressed NRR only above the Schwarzschild barrier \citep{Alexander2017}; thus we do not expect our results to be affected by resonant relaxation.

{ Finally, our orbital treatment is limited to a purely Newtonian description. }

\section{Results}\label{sec:results3}

\begin{figure}
\centering
 \vspace*{-5mm}
\includegraphics[width=.44\textwidth]{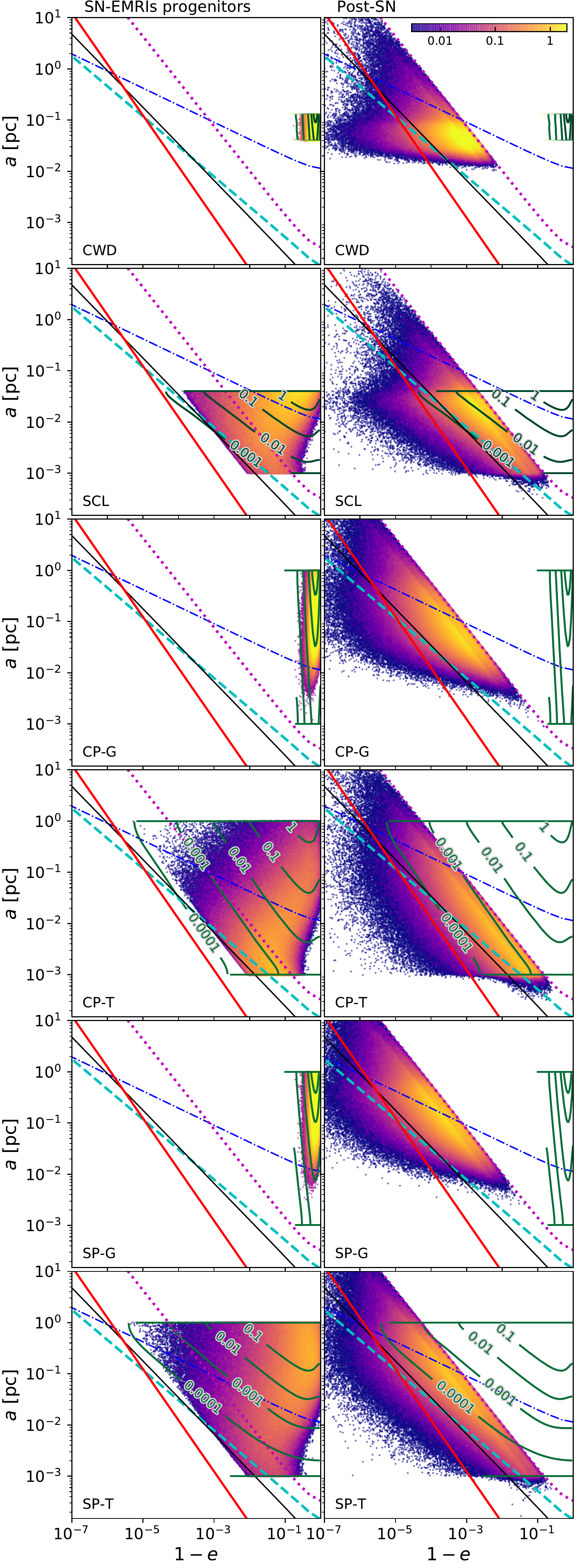}
 \vspace*{-2mm}

\caption{\small  Distribution of stars and COs  in the $(1-e, a)$ plane prior and after the `fast' SN-kick, for each of the initial young stellar structures. The lines are the same  as presented in Fig.~\ref{fig:fig123}. In all panels, dark-green isocontours indicate how the population of progenitor stars is initially distributed, while the dark-green horizontal lines indicate their minimum and maximum allowed $a_i$. The left-hand scatter plots show the initial position of those progenitor stars which end up as SN-EMRIs in the right-hand plot (assuming the H population); the right-hand scatter plots show how COs redistribute after the `fast' kick (for convenience, we only display COs for which $t_{\rm GW}<14$ Gyr). }
\label{fig:scatter}
\end{figure}

\begin{table*}
  \centering
  \caption{Statistics of SN-EMRIs, direct plunges, unbound objects, tidal disruption events and premature GW-inspirals for each of the different young stellar populations assumed to inhabit the GC}
  \label{tab:emris}
  \begin{center}
  
\begin{tabular}{lcccccc} \hline
Population  & EMRIs, H & EMRIs, L & direct plunges & unbound & TDE & pre-GW \\ \hline
CWD & $1.4\times 10^{-5}$ ($6.8\times 10^{-6}$) & $8.1\times 10^{-6}$ ($4.1\times 10^{-6}$) & $9.9\times 10^{-6}$ ($4.9\times 10^{-6}$) & $0.396$ ($0.199$) & $<10^{-10}$  & $<10^{-10}$\\
SCL & $3.2\times 10^{-4}$ ($2.7\times 10^{-4}$) & $4.2\times 10^{-4}$ ($3.7\times 10^{-4}$) & $9.8\times 10^{-5}$ ($9.0\times 10^{-5}$) & $0.118$ ($0.059$) & $4.6\times10^{-4}$ & $2.8\times10^{-4}$\\
CP-G & $2.4\times 10^{-6}$ ($1.2\times 10^{-6}$) & $1.5\times 10^{-6}$ ($7.3\times 10^{-7}$) & $2.2\times 10^{-6}$ ($1.2\times 10^{-6}$) & $0.750$ ($0.408$) & $<10^{-10}$  & $<10^{-10}$\\
CP-T & $3.7\times 10^{-5}$ ($2.9\times 10^{-5}$) & $6.4\times 10^{-5}$ ($5.6\times 10^{-5}$) & $1.0\times 10^{-5}$ ($1.0\times 10^{-5}$) & $0.760$ ($0.411$) & $6.2\times10^{-5}$ & $3.2\times10^{-5}$\\
SP-G & $6.6 \times 10^{-7}$ ($3.4 \times 10^{-7}$) & $2.8 \times 10^{-7}$ ($1.4 \times 10^{-7}$) & $1.0 \times 10^{-6}$ ($6.4 \times 10^{-7}$) & $0.836$ ($0.462$) & $<10^{-10}$  & $<10^{-10}$\\
SP-T & $2.2 \times 10^{-6}$ ($2.1 \times 10^{-6}$) & $2.2 \times 10^{-6}$ ($1.9 \times 10^{-6}$) & $1.6 \times 10^{-6}$ ($3.2 \times 10^{-6}$) & $0.847$ ($0.464$) & $2.1\times10^{-5}$& $3.2\times10^{-6}$\\
 \hline
\end{tabular}

  \end{center}
\justifying
{\footnotesize First column:  different young stellar populations assumed to inhabit the GC (Tab.~\ref{tab:ic3}); second (third) column: fraction of SN kicks resulting in the generation of  EMRIs, assuming the H (L) population to dominate NRR effects; fourth column: fraction of SN events resulting in a direct plunge; fifth column: fraction of SN explosions unbinding the newborn CO from the SMBH; sixth column: fraction of stars undergoing a tidal disruption event prior to the SN explosion; last column: fraction of objects (excluding those reported in the previous column) for which the progenitor stars $t_{\rm GW}$ is shorter than the time they need to undergo SN: most probably, these objects will undergo a tidal disruption resulting from their GW-induced decay prior to the SN explosion. Non-bracketed numbers refer to the statistics obtained assuming `fast' BH kicks, while numbers in parenthesis assume the `slow' BH kicks; the statistics in the last two columns is not affected by the magnitude of the kick. 
}
\end{table*}

The total fractions of SN kicks resulting in the generation of EMRIs   for each of the  assumed  populations  are listed in Tab.~\ref{tab:emris}, while Fig.~\ref{fig:scatter} shows how COs distribute in the ($a$, $1-e$) phase space prior and after the SN-kick { (in particular, the right hand scatter plots only show COs for which $t_{\rm GW}$ is shorter than the Hubble time)}. As we will better detail below, a star undergoing a SN is more likely to induce an EMRI if it is initially located in the vicinity of (or inside) the `EMRI' region of phase-space. 
In other words, SN progenitors at small separations from the SMBH and/or spanning an initially eccentric orbit are more likely to induce EMRIs.

\subsection{EMRI and plunge statistics}

Tab.~\ref{tab:emris} shows that the fraction of SN explosions that produce EMRIs lies in the range $10^{-7} - 10^{-4}$. The S-cluster represents  the most efficient structure for the production of SN-EMRIs  ($\sim 4\times10^{-4}$), owing to its small distance from the SMBH ($a_i<0.04$) and to its TE distribution. In the CW disc, the fraction of SNe triggering EMRIs is roughly one order of magnitude lower; this is mostly due to the larger distance of the CW disc from the SMBH and to the low orbital eccentricity of progenitor stars within the CW disc. However, the CW disc appears to host a substantially larger number of stars than  the S-cluster, thus the overall amount of SN-EMRIs triggered in the CW disc could still be larger than the amount of events expected in the S-cluster.

The concentrated profile (CP) is much more efficient in generating SN-EMRIs compared to the shallow one (SP), as many more objects are available at small separations from the SMBH;  the initial eccentricity is also important: the TE cases result in at least an order of magnitude greater fraction of SN-EMRIs compared to the GE cases. 
 Finally, different assumed background populations (H, L) producing NRR only marginally influence  the relative fraction of SNe triggering EMRIs. 

A larger amount of SN-EMRIs are induced in the fast BH kick assumption compared to the slow kick scenario. Larger kicks have a more significant impact on the initial orbit: even if they unbind roughly two times more COs (as shown in the fifth column of Tab.~\ref{tab:emris}), they still manage to trigger more SN-EMRIs, as the GW events typically originate from progenitors with $a_i\lesssim 0.1$ pc (see the scatter plots in the left-hand panels of Fig.~\ref{fig:fig123}), where COs hardly get unbound by the kick.
On the other hand, the difference between fast kicks and slow kicks in terms of EMRI's fraction is not dramatic: both the fast and slow kick assumptions lead to the same order of magnitude of SN-EMRIs.

\subsection{EMRI rates as a function of the initial conditions}

\begin{figure}
\centering
\includegraphics[width=0.35\textwidth]{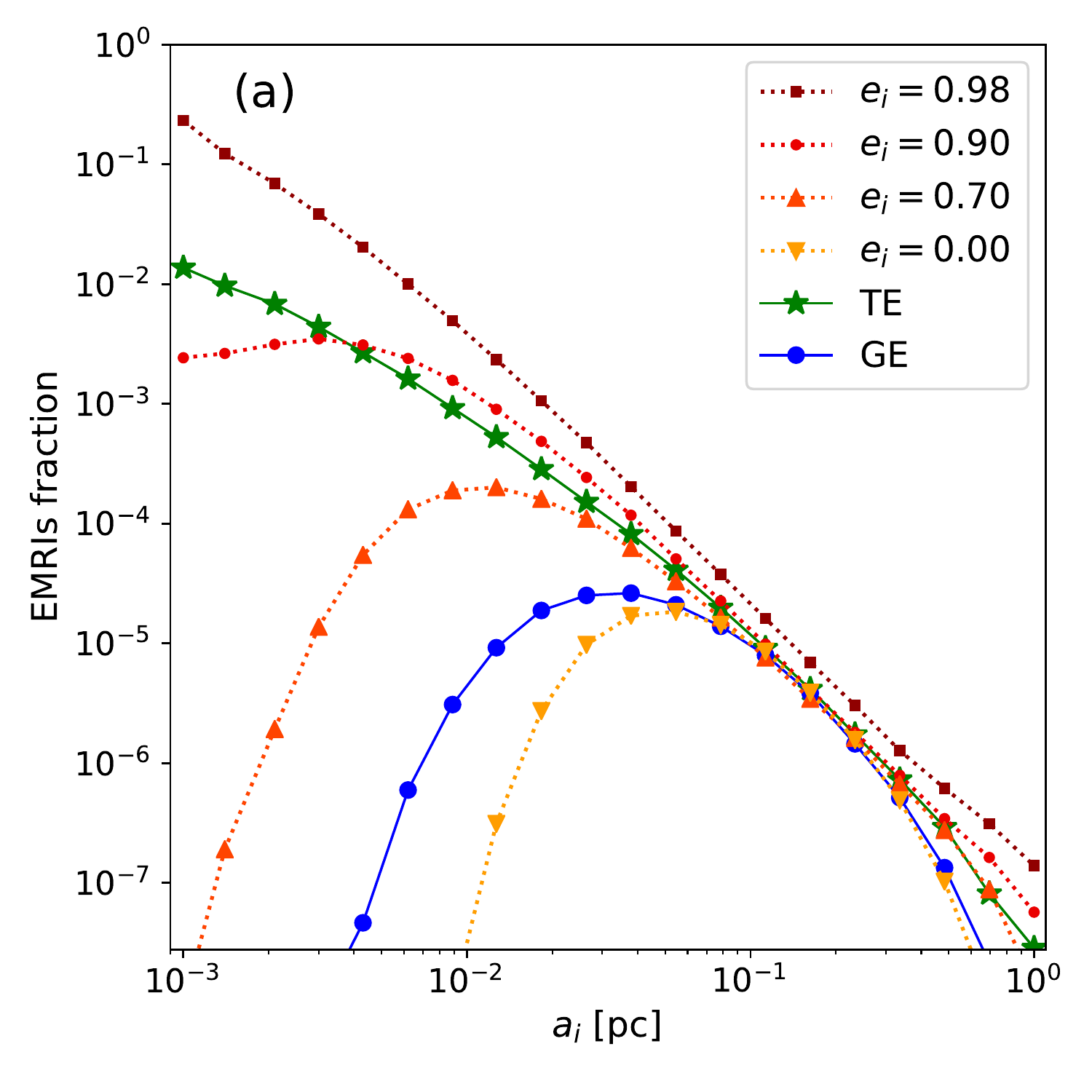} 
\includegraphics[width=0.35\textwidth]{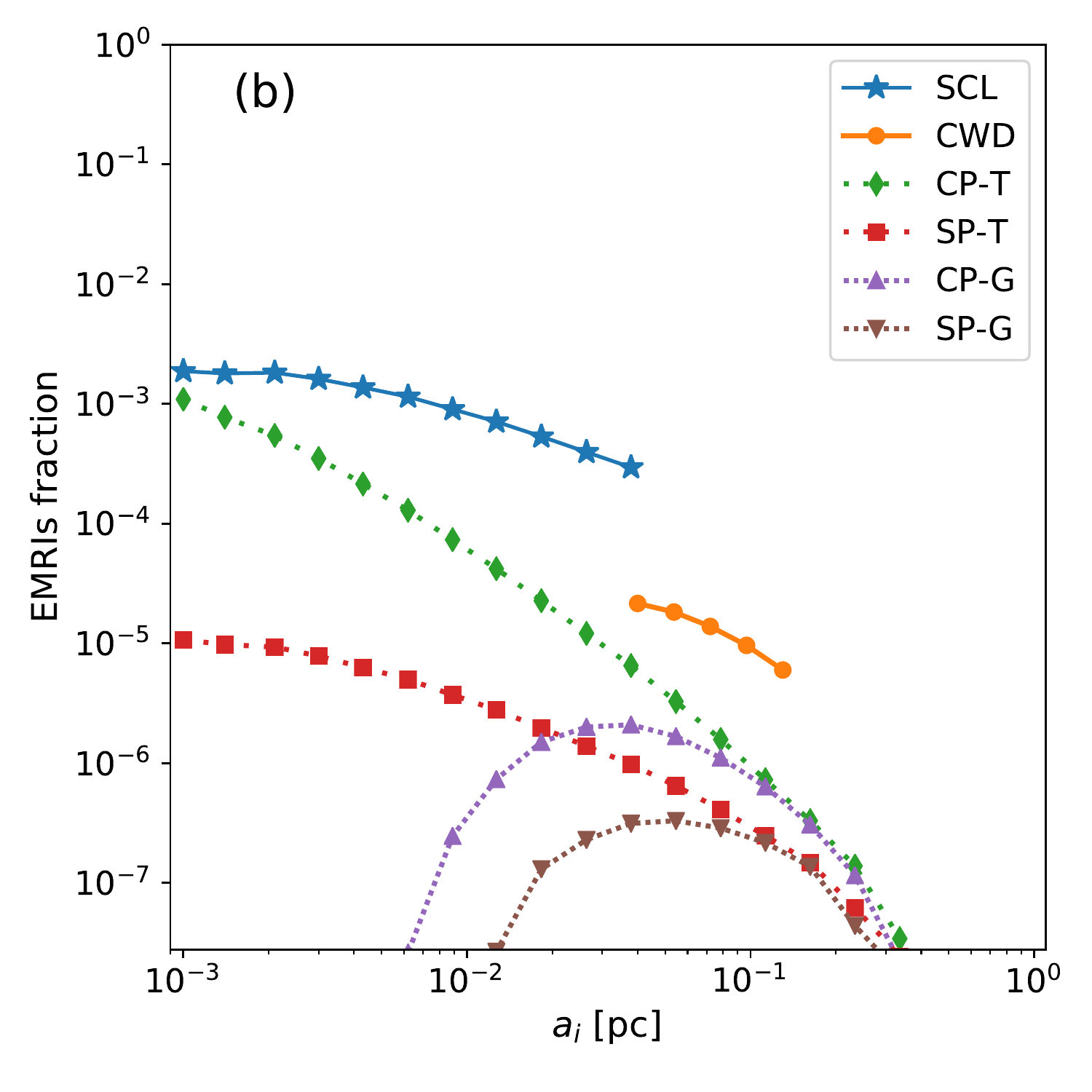} 
    \caption{Fraction of stars undergoing a SN that become EMRIs, as a function of the initial semi-major axis of the progenitor star, $a_i$. In (a), each point assumes a Dirac delta distribution with the same height for each $a_i$ (i.e., no particular density distribution for the progenitor stars is considered); different curves in (a) show different values for the initial eccentricity: $e_i$ is either fixed at $0, 0.7,0.9,0.98$ (dotted lines), or a TE   (green stars) and a GE distribution (blue circles) is considered. Panel (b)  shows  the probability for a SN occurring  within $a_i, a_i+{\rm d}a_i$ to result in a SN-EMRI, assuming the semi-major axis and eccentricity distributions in Tab.~\ref{tab:ic3}. We stress that the  curves for the S-stars and CW disc attain larger values as  these systems are far less extended than the other structures investigated, thus a larger fraction of progenitor stars is allowed within each ${\rm d}a_i$. Here, we assumed the `fast' kick prescriptions and the `H' background population; if the `slow' kick prescription and/or the L background population is considered, the displayed curves behave similarly.  }
    \label{fig:femris_a}
\end{figure}

In order to better interpret the relative fraction of SNe inducing EMRIs in Tab.~\ref{tab:emris}, here we analyse the probability that a progenitor with a given initial $a_i$ and $e_i$ (or following a given distribution of these two parameters) will trigger an EMRI via its SN kick. The fraction of SNe resulting in SN-EMRIs as a function of the progenitor star semi-major axis is shown in Fig.~\ref{fig:femris_a}a: each point in the plot assumes a Dirac delta distribution for $a_i$, and $e_i$; we further explored the TE and GE distributions for each $a_i$.
Three important regimes can be distinguished in the plot.
\begin{enumerate}

\item At the smallest considered scales ($a_i\sim 10^{-3}$ pc), only progenitor stars with high initial eccentricity ($e_i\gtrsim 0.9$) induce a substantial fraction of SN-EMRIs ($\sim 0.001$ to $\sim 0.1$); progenitor stars with lower orbital eccentricities ($e_i\lesssim0.7$), instead,  trigger a negligible amount of GW events. In fact, the typical stellar orbital velocities about the SMBH at these small separations are extremely high ($\gtrsim 1500$ km s$^{-1}$) compared to the average kick velocity  ($v_k\approx 400$ km s$^{-1}$ in the `fast' BH kick scenario assumed in Fig.~\ref{fig:femris_a}a). As a result,  SN-EMRIs can be induced at such small distance from the SMBH only if a very small perturbation to the progenitor orbit suffices to funnel the CO in the `EMRI' area of phase space: progenitors inducing SN-EMRIs at $a_i\sim 10^{-3}$ pc are thus either already inhabiting, or  very close to the `EMRI' area; lower eccentricity parent stars, instead, would require much larger kicks to induce the GW events.

\item At intermediate scales ($10^{-2}\lesssim a_i/{\rm pc}\lesssim 10^{-1}$), SN kicks occurring in the GC have about the same speed as stars orbiting the SMBH. It follows that SN kicks in this region have the best chance to significantly perturb primitive stellar orbits without unbinding a large fraction of COs. The plot shows that a non-negligible amount of  SN-EMRIs can be produced regardless of the assumed initial eccentricity, even in the    $e_i=0$ case.  

\item At the largest investigated scales  ($a_i\gtrsim 0.1$ pc) the average velocity kick imparted to a CO exceeds the typical velocity required to unbind it from the SMBH (Fig.~\ref{fig:prob_e}). For this reason,  most COs with $a_i>0.1$ pc get unbound from the SMBH and cannot undergo an EMRI.

\end{enumerate} 

Fig.~\ref{fig:femris_a}a also shows that the GE curve behaves as the curves describing the low, fixed eccentricity cases; in fact, the mean GE eccentricity is $\approx0.3$. On the other hand, the TE curve shows a significant fraction of SN-EMRIs at $a_i<10^{-2}$ pc, in spite of the fact that its mean eccentricity is $\approx0.7$: this is because a significant number of objects have large eccentricities in the  TE distribution, as 19\% (2\%) of objects attain $e_i>0.9 \ (0.99)$.

Fig.~\ref{fig:femris_a}a does not give a realistic physical estimate of the fraction of SNe resulting in EMRIs  at different radii, as no physically motivated distribution is assumed for the semi-major axes of the progenitor stars. Fig.~\ref{fig:femris_a}b, instead, shows the probability for a star whose initial semi-major axis lies in the range ($a_i,a_i+{\rm d}a_i$) to become a SN-EMRI,  assuming the progenitor stars semi-major axis and eccentricity distributions presented in Tab.~\ref{tab:ic3}.
As already mentioned, more concentrated stellar distributions favour the generation of EMRIs close to the SMBH, as more objects are available at the smallest scales  where (i) SNe hardly unbind newborn COs from the SMBH and (ii) progenitors  are generally already close to the `EMRI' phase space area. 
Fig.~\ref{fig:femris_a}b   shows that, assuming the same $a_i$ distribution,   TE cases trigger a significantly larger amount of SN-EMRIs, compared to GE cases. The reason has  to be sought in the low fraction of progenitor stars that  are initially placed near the low-angular momentum, EMRI region of phase space in the GE assumption.

\subsection{Direct plunges}
Table~\ref{tab:emris} shows that a fraction $\sim 10^{-6}- 10^{-4}$ of SN kicks result in a direct plunge. This fraction is typically of the same order of magnitude as the fraction of SN-induced EMRIs for each of the investigated populations. 

In general, the relative fraction of direct plunges to SN-EMRIs is larger when the initial conditions allow a larger amount of objects to be available at large scales: in fact, only direct plunges can occur if the final semi-major axis exceeds a few $\times 0.1$ pc (Fig.~\ref{fig:fig123}). This means that the ratio between number of plunges and number of EMRIs is higher if the initial conditions extend to larger separations from the SMBH, and if the initial density profile of progenitors is shallower.
This finding is in agreement with the studies that estimate the ratio of plunges to EMRIs induced by standard NRR processes \citep[e.g.][]{Merritt2015,Babak2017}.

\subsection{Lost progenitor stars}

Table~\ref{tab:emris} further reports the statistics of progenitor stars that are `lost' prior to undergo a SN because they get disrupted by the SMBH tidal field. In particular, we distinguish between progenitors undergoing a `standard' tidal disruption event solely due to the fact that their initial pericentre is too close to the SMBH, and those that decay onto the SMBH via GW emission (thus being again tidally disrupted) prior to experience a SN. No progenitor is lost in the GE cases, as the parent stars attain quite large pericentre distances prior to the SN; in the TE cases, instead, the fraction of objects experiencing a standard tidal disruption is $1-10$ times the listed fraction of SN-EMRIs. Progenitors inspiralling onto the SMBH via GW radiation prior to the SN are typically half to one tenth of those undergoing the standard tidal disruption.

\subsection{Progenitors in the `EMRI' region}


\begin{table*}
  \centering
  \caption{Fraction of objects in the EMRI region prior to the SN}
  \label{tab:from0}
  \begin{center}
  
\begin{tabular}{lcccccc}
\hline
\multirow{2}{*}{Structure} & \multicolumn{2}{l}{ {Already in `EMRI' region} } & \multicolumn{2}{l}{SN-EMRIs from `EMRI'} & \multicolumn{2}{l}{COs staying in `EMRI'} \\
 & H & L & H & L & H & L \\ \cmidrule(lr){1-1} \cmidrule(lr){2-3} \cmidrule(lr){4-5}\cmidrule(lr){6-7}
CWD & 0  & 0  &  -- (--) &  -- (--) &  -- (--) & -- (--) \\
SCL & $1 \times 10^{-5}$  & $1 \times 10^{-4}$  & 1\% (1\%) & 8\% (15\%) & 24\% (28\%) & 29\% (47\%) \\
CP-T & $4 \times 10^{-6}$  & $3 \times 10^{-5}$  & 3\% (4\%) & 20\% (35\%) & 29\% (31\%) & 37\% (55\%) \\
SP-T & $6 \times 10^{-8}$ & $6 \times 10^{-7}$ & \multicolumn{1}{l}{1\% (1\%)} & 8\% (14\%) & 24\% (30\%) & 31\% (48\%) \\ \hline
\end{tabular}

  \end{center}
\justifying
{\footnotesize First column: young stellar populations assumed to inhabit the GC; second and third column: fraction of the total SN progenitors already inhabiting the `EMRI' region of phase space  prior to the SN explosion; fourth and fifth column: fraction of  SN-EMRIs from stars already in the `EMRI' phase space region prior to the SN, relative to the total amount of objects undergoing SN-EMRIs;  second-to-last and last column: fraction of objects among those inhabiting the `EMRI' region prior to the SN that keep sitting in the same region after the kick, thus undergo a SN-EMRI. The `H' and `L' indicate that we assumed respectively the H or L background population to dominate NRR effects. Non-bracketed numbers refer to the statistics obtained assuming `fast' BH kicks, while numbers in parenthesis assume the `slow' BH kicks when the two different choices make a difference. The statistics is obtained by performing $\sim10^{10}$ Monte-Carlo experiments for each population; we do not show the GE cases (apart from the CW disc one) as their fraction of stars in the EMRI region prior to the SN is <$10^{-10}$ of the total.   }
\end{table*}

Figure~\ref{fig:scatter} shows that a certain amount  of progenitor stars  inhabit the `EMRI' region  prior to the SN kick, and a few of them remain in the same region after the kick: we report the statistics associated to this population in  
Tab.~\ref{tab:from0}. 

A small amount ($<10^{-4}$)  of  progenitor stars are  in the EMRI region prior to the SN kicks; the fraction of SN-EMRIs resulting from this `lucky' sub-sample is $\lesssim{}35\%$ of al SN-EMRIs, depending on the background and SN kick assumptions. Therefore, the bulk of SN-EMRIs are \textit{not} triggered by progenitor stars inhabiting the `EMRI' area prior to the kick. In particular, we stress that \textit{no} SN-EMRIs from the CWD distribution were already in the `EMRI' region before the SN explosion. 

Furthermore,  $\approx25\%$ to 60\% of progenitors initially in the EMRI area keep staying there and become SN-EMRIs after the kick. This means that stars that form  on an EMRI trajectory have a significant probability of undergoing an EMRI even if they experience a SN kick, and this is true even when the fast BH kick prescriptions are adopted.

\subsection{SN-EMRI mass function}

\begin{figure}
\centering
\includegraphics[ trim={6mm 0mm 18mm 10mm},clip,width=0.43\textwidth]{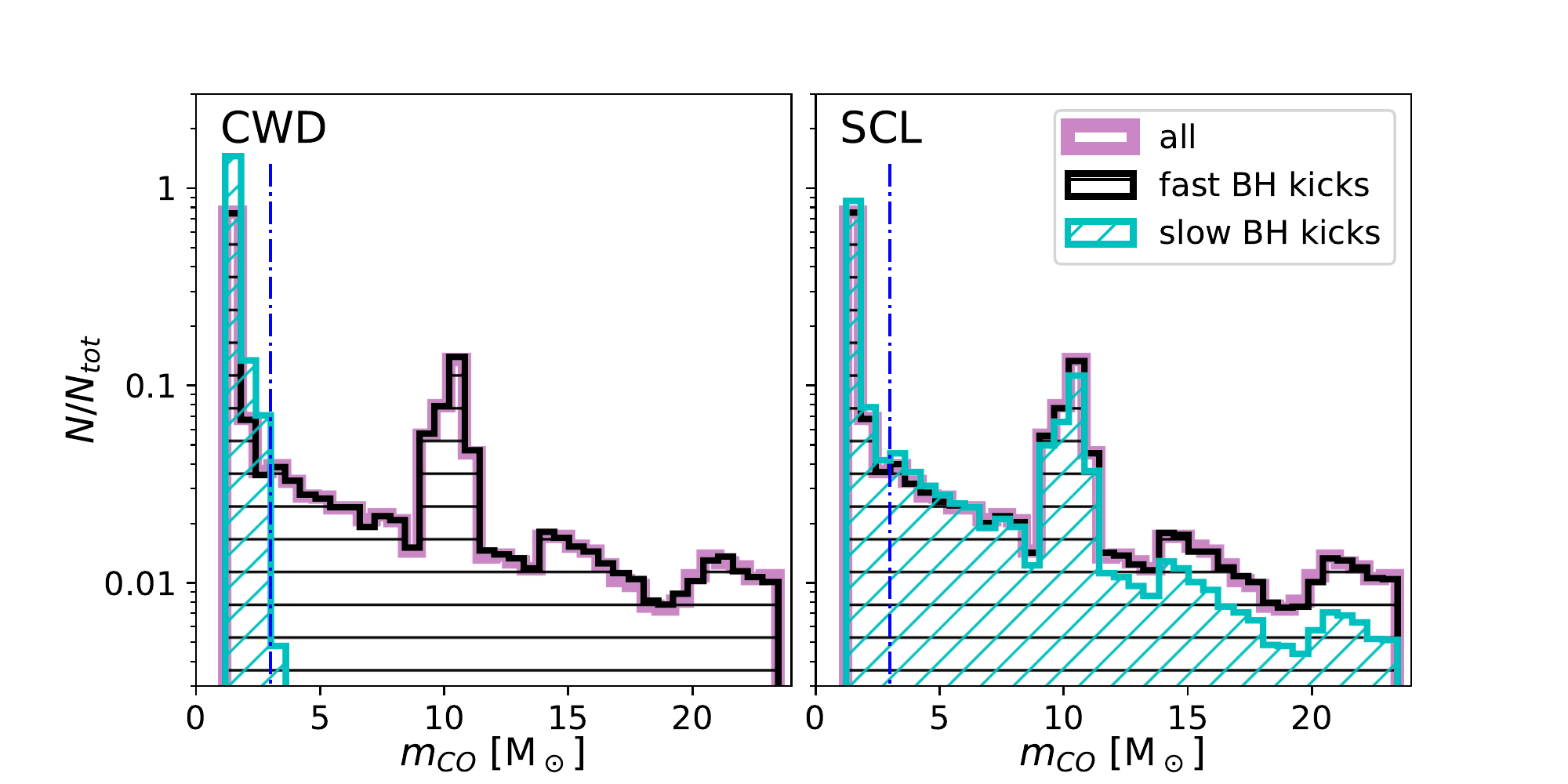}
    \caption{Distribution of CO masses. The thick violet line shows the underlying distribution obtained from the initial top-heavy mass function; the black and cyan histograms show the mass distribution of SN-EMRIs undergoing a fast or slow BH kick, respectively. We show the CWD case (left) and the SCL case (right). All the shown distributions are normalized to unity; the vertical dash-dotted line delimits what we define to be NSs ($<3 \msun$) and BHs ($\geq 3 \msun$). }%
    \label{fig:mfunc3}
\end{figure}

The assumed top-heavy mass function produces an even relative fraction of NSs and BHs (0.497 and 0.503, respectively), while the fraction of COs exceeding 10 $\msun$ is 0.29; here we compare this underlying CO mass distribution to the SN-EMRI one. 

If the fast BH kick prescriptions are adopted, the mass function of SN-EMRIs is largely consistent with the primitive one in all considered scenarios (see Fig.~\ref{fig:mfunc3} for the CWD and SCL cases), as the $a_{\rm NRR}$ separatrix (Eq.~\ref{eq:aNRR}) does not depend on the CO mass. 

In contrast, if the slow BH kick prescriptions are assumed,  SN-kicks are far less efficient in scattering the heaviest COs in phase space. As a consequence, the relative fraction of massive BHs undergoing SN-EMRIs decreases. The drop is more consistent if the GE distribution is assumed (e.g. in the CWD case), as shown in Fig.~\ref{fig:mfunc3}. In fact, when the slow BH kicks are adopted, $40\%$ (<$1\%$) of SN-EMRIs are BHs if the progenitors follow a TE (GE) distribution.

\subsection{Inspiral time-scales}

\begin{figure}
\centering
\includegraphics[width=.43\textwidth]{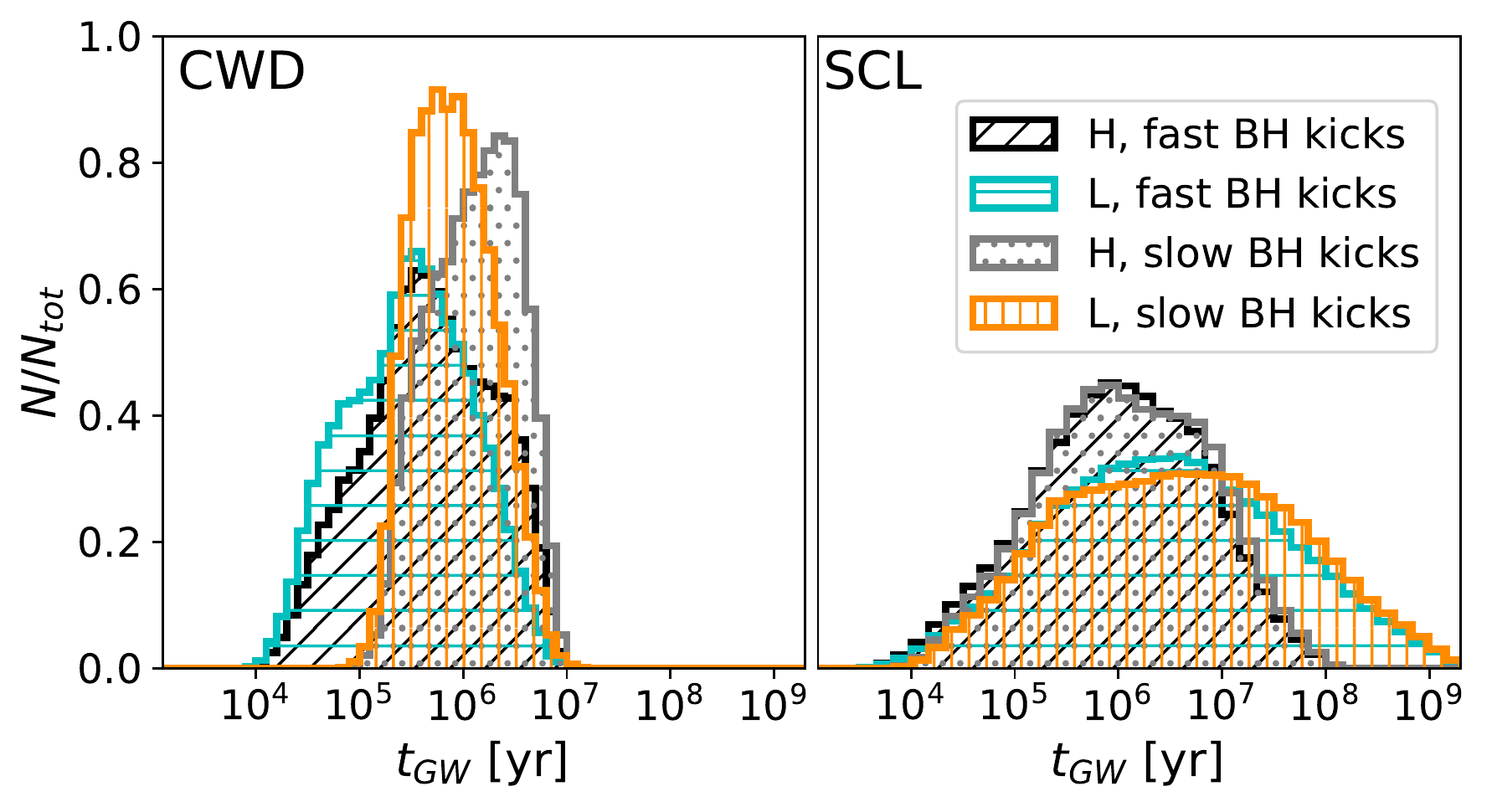}
    \caption{Distribution of the time elapsed from the SN explosion to the final CO coalescence {(computed via Eq.~\ref{eq:tgw_via_e})}, for the COs undergoing a SN-EMRI. We show only the CWD (left) and SCL (right) scenarios, as they are representative of the GE and TE cases, respectively. Within each panel, the black and grey (cyan and orange) histograms assume the H (L)  background population and the fast and slow BH kick prescriptions, respectively.}%
    \label{fig:tinfall}
\end{figure}

Fig.~\ref{fig:tinfall} shows the distribution of time elapsed from the SN explosion to the final merger into the SMBH, accounting solely for the COs undergoing a SN-EMRI for two representative cases. 
{ 
The inspiral timescale here is computed  via

\begin{subequations}
\label{eq:tgw_via_e}
\begin{eqnarray}
t_{{\rm GW},e} =& -\frac{12}{19}\frac{c_0^4}{\beta} \int_e^0 \frac{\Tilde{e}^{29/19}}{(1-\Tilde{e}^2)^{3/2}}  { \left(1+\frac{121}{304}\Tilde{e}^2\right)^{1181/2299}}d\Tilde{e} ;\\
c_0 =& \frac{a(1-e^2)}{e^{12/19}\left[1+\left(121/304\right)e^2\right]^{870/2299}};\\
\beta =& \frac{64}{5}\frac{G^3 M M_\bullet m_{\rm CO} }{c^5},
\end{eqnarray}
\end{subequations}
following \citet{Peters1964}. Eq.~\ref{eq:tgw_via_e} computes the time required by GWs to circularize the CO orbit, and it provides a more accurate estimate of the inspiral timescale compared to Eq.~\ref{eq:tgw3};
the two approaches typically lead to GW-inspiral timescales that are compatible within a factor $\lesssim 2$.
}
Fig.~\ref{fig:tinfall} suggests that most objects complete their inspiral in $10^{5 -7}$ yr.  

The inspiral time distributions exhibit a  different shape depending on the choice of initial conditions. The S-stars, representative of the TE case,   display  a broader distribution of times compared to the CW disc (which is representative of the GE cases), especially if the L background population induces NRR.

The typical inspiral times reported here are compatible to what found by \citet{Lu2018}, addressing SN kicks in binary stars about a Milky-Way-like SMBH. \citet{Lu2018} also suggest GW events can occur over a time-scale as short as a few minutes after the SN, but they do not distinguish between proper EMRIs and direct plunges.



\subsection{Energy and angular momentum conservation}\label{sec:alaytic}

\begin{figure}
\centering
\includegraphics[ trim={2mm 1mm 0mm 12mm},clip,width=.43\textwidth]{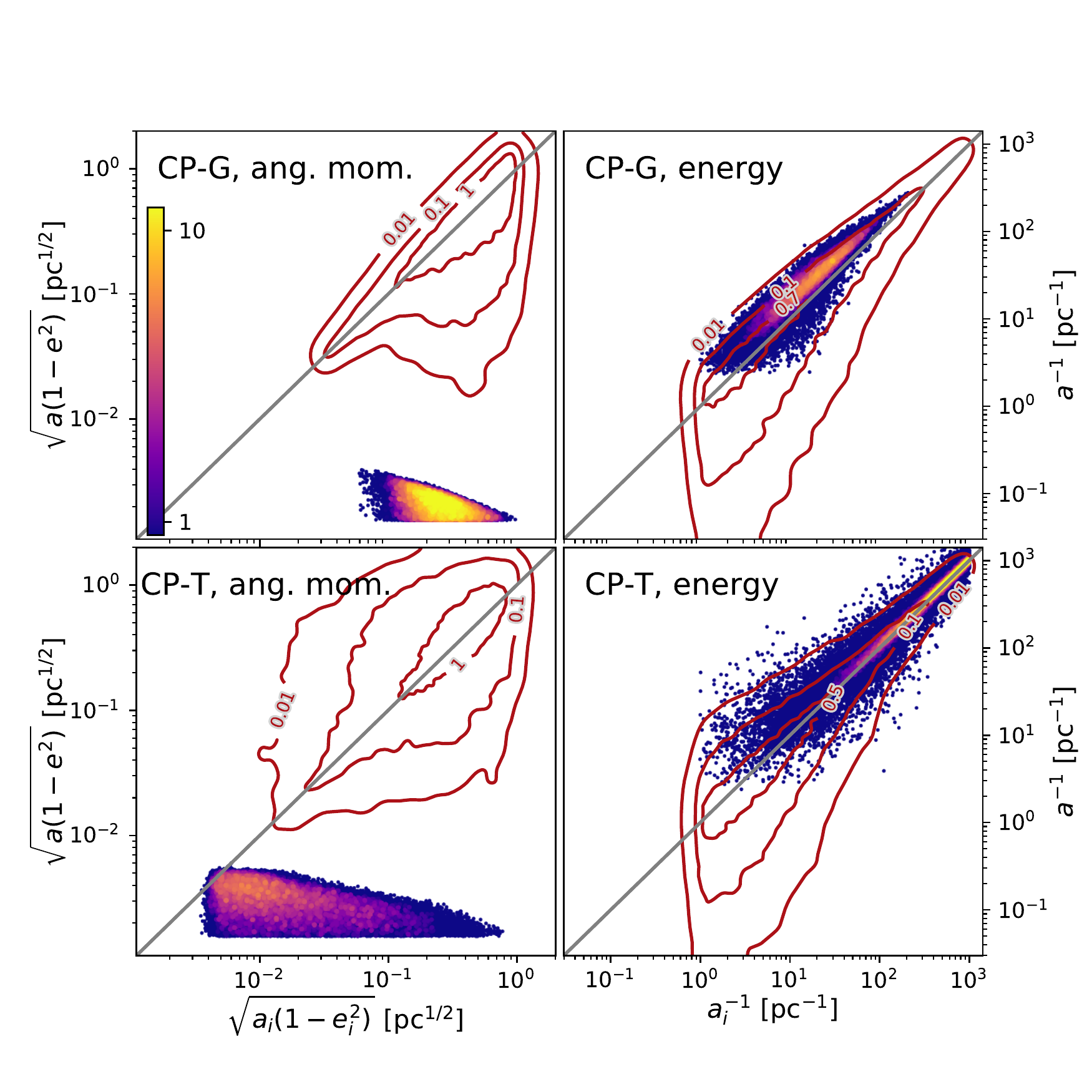}
    \caption{The two left-hand plots show how COs redistribute in the angular momentum space: we display the initial versus the final angular momentum magnitude divided by $\sqrt{G\,{}M}$ for the CP-G (top) and CP-T (bottom) case. The two right-hand panels show the initial versus the final inverse semi-major axis, which is a proxy for the  orbital binding energy, again  for the CP-G (top) and CP-T (bottom) case. In all displayed panels, scatter plots refer to the sole objects undergoing SN-EMRI (in the fast BH kick and H background assumption), while the contours show the distribution of all objects that remain bound after the kick.  }%
    \label{fig:je_scatter}
\centering
\includegraphics[ trim={1mm 0mm 8mm 5mm},clip,width=.43\textwidth]{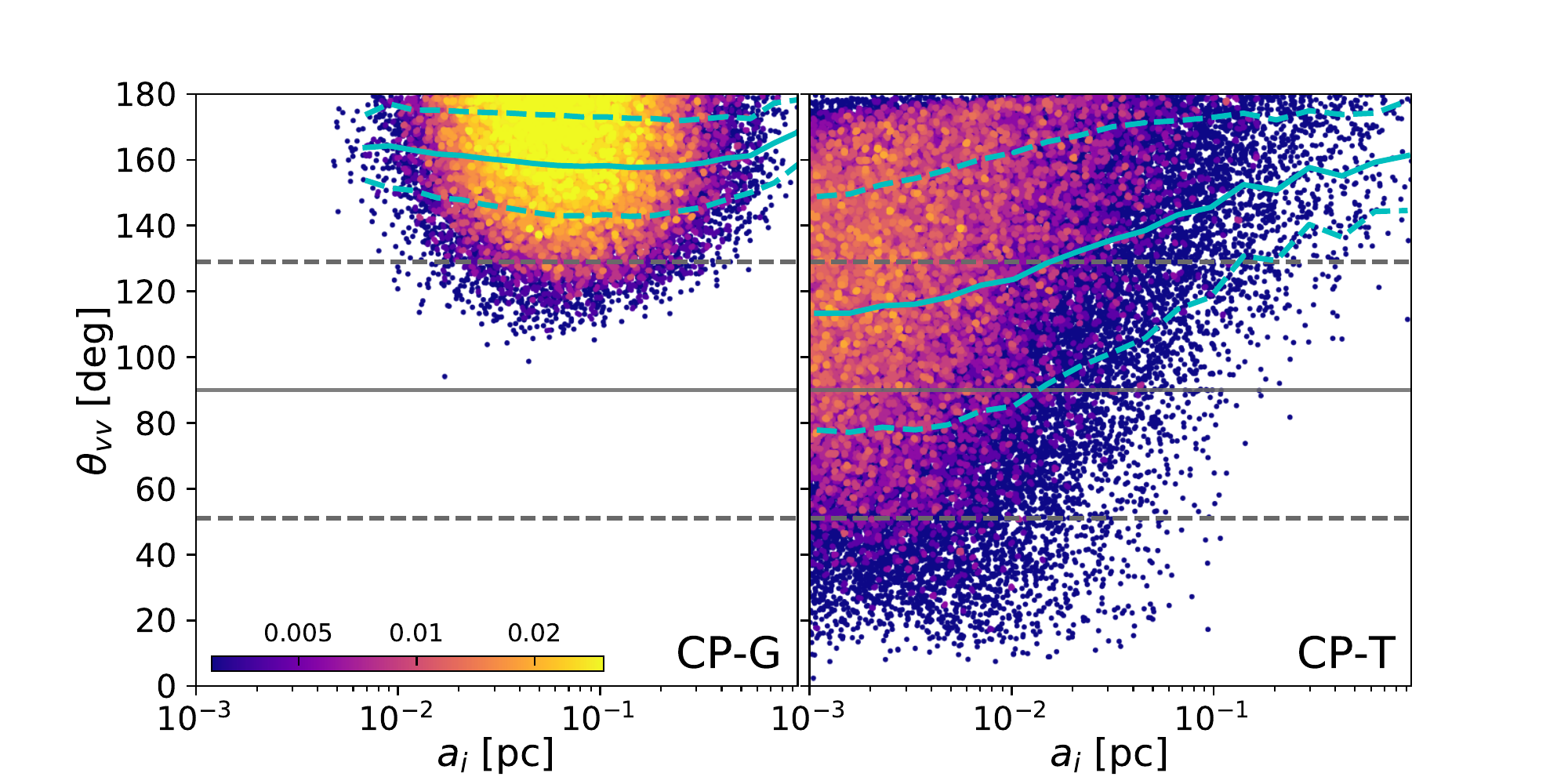}
    \caption{Angle between the initial orbital velocity vector at the moment of the SN explosion  and the vector of the SN kick ($\theta_{\rm vv}$), as a function of the initial semi-major axis, for the objects undergoing a SN-EMRI. We show the cases assuming a CP profile with the GE (TE) distribution in the left-hand (right-hand) plot. In both panels, the cyan solid line shows the mean value of $\theta_{\rm vv}$ as a function of $a_i$, while the cyan dashed lines show the mean $\pm$ one standard deviation. The horizontal black solid line at $90^\circ$ shows the mean $\theta_{\rm vv}$ we would expect if it were isotropically distributed, while the dashed lines show the mean value $\pm$ one standard deviation. We show the fast BH kicks and the H background population. }%
    \label{fig:thetavv}
\end{figure}

Figure~\ref{fig:je_scatter} shows the initial versus the final specific binding energy and specific angular momentum magnitude of SN-EMRIs (data points) and of all the COs that remain bound to the SMBH (contours) in two representative cases.

On average, the whole collection of COs that keep orbiting the SMBH tend to maintain the same specific binding energy after the kick. The same can be said  for most objects that undergo SN-EMRIs: that is, most COs undergoing SN-EMRIs maintain the same energy, on average, prior and after the kick.

The overall distribution in angular momentum magnitude of all objects bound to the SMBH is again not significantly affected by the kick. On the other hand, SN-kicks triggering successful SN-EMRIs attain  significantly lower final angular momentum magnitudes than the rest of COs. In fact, EMRIs typically require very low eccentricities to be produced. Thus, we can state that SN-EMRIs are produced if the SN kick almost erases the orbital angular momentum of the progenitor star.

Figure~\ref{fig:thetavv} shows  the angle between the original Keplerian velocity of the progenitor and the SN velocity kick distribution, as a function of the initial semi-major axis of the progenitor, for two representative cases. We only display the COs that undergo a SN-EMRI. 
The plots confirm that SN explosions efficiently trigger EMRIs if the angle between the kick and the initial velocity almost cancel out, so that the final object finds itself onto a very low angular momentum orbit. 

\section{Summary and discussion}\label{sec:disc_concl}

We used a Monte-Carlo method to address the possibility that natal kicks of NSs and BHs could trigger EMRIs in the GC. We showed that $10^{-7}$ to $10^{-4}$ of SN events occurring in a GC-like environment are expected to trigger an EMRI. We call these events SN-EMRIs. 
Below we summarize and discuss our key results.

\begin{itemize}

\item 
SN-EMRIs are produced much more efficiently if the stellar orbits have large eccentricity, e.g. if their eccentricities distribute thermally. However, even initial orbital eccentricities $\lesssim{}0.4$ result in the generation of $10^{-7}-10^{-5}$ EMRIs per SN.
The majority of SN-EMRIs are associated with an initial orbital semi-major axis $<0.1$ pc,  and typically  of the order of $10^{-2}-10^{-3}$ pc.

\item 
Our investigation shows that S-stars are located in an optimal region of the phase space for triggering SN-EMRIs: $3-4\times10^{-4}$ of SNe occurring in this region lead to the  production of SN-EMRIs.  
SN kicks occurring in the CW disc have a lower but still significant probability of generating SN-EMRIs:  approximately one in $10^5$. Since the CW disc is populated by more stars compared to the S-cluster, the former could still induce a higher rate of SN-EMRIs compared to the S-cluster.

\item In this paper we only focused on SN kicks that directly induce an EMRI event. Plausibly, a number of COs could get closer to the EMRI region of phase space as a result of their natal kick (although not entering it); these objects would have a significant chance to be pushed into the EMRI region by NRR (Re'em Sahari, private communication). In this sense, our estimated rate only represents a lower limit to the number of EMRIs actually triggered by SNe.

\item  SN kicks induce approximately the same order of magnitude of EMRIs and direct plunges. If the central SMBH has a large spin parameter, though, a fraction of events classified as direct plunges in this study could be real EMRIs in disguise, as the SMBH ISCO moves closer to a rapidly spinning SMBH if the CO inspirals on a prograde orbit \citep{Amaro-Seoane2013}. { In addition, part of the events that have been labeled as direct plunges in this study  might in fact contribute to the LISA signal as extreme-mass-ratio-bursts \citep{Rubbo2006}, i.e. signals from very eccentric COs that emit a detectable burst of GWs only at periapsis. This signal can be detected by LISA only if it originates from the GC or a nearby galactic nucleus; since many periapsis passages may occur per each of such very eccentric inspirals, these events could still lead to a moderate detection rate for LISA \citep{Hopman2007,Berry2013}. }

\item Even if the bulk of SN-EMRIs is scattered into the EMRI area from other regions of the phase space, a few progenitors inhabit the EMRI region from the beginning ($10^{-4}-10^{-7}$,  depending on the initial distribution of $a_i$ and $e_i$). This population of progenitors has a consistent chance of remaining in the EMRI region of phase space even after the kick ($30-50\%$). This fact is relevant as some alternative EMRI production mechanisms include  (i) the in-situ formation of COs within the SMBH accretion disk \citep{Levin2003},  and (ii) the capture by the disc of massive stars (later turning into COs) on orbits that cross it %
\citep[e.g.][]{Syer1991,Rauch1995,Panamarev2018}. Our study suggests that such objects have a significant chance to remain on an EMRI orbit even after receiving a quite large natal kick; thus, gas drag and slow natal kicks have not necessarily to be invoked in order for a successful EMRI to be generated.

\item In general, the derived distributions of $t_{\rm GW}$ (the time elapsed from the SN kick to the final plunge onto the SMBH ISCO) clearly suggests that most SN-EMRIs complete their inspiral in $10^{5-6}$ yr. This aspect is remarkable as the aforementioned time-scales roughly correspond  to  the age of the observed young population at the GC \citep{Lu2013,Habibi2017} and are also of the same order of magnitude as the time elapsed from the birth of a massive star to its SN explosion. It follows that the most massive progenitors born in the most recent star formation episode at the GC may have undergone their SN a few Myr ago, and they might be on their route to undergo a SN-EMRI. 
 
 The distribution of $t_{\rm GW}$ further suggests that, in galactic nuclei similar to the Milky Way, SN-EMRIs could be detected in coincidence with a still undergoing star formation episode close to the SMBH. On the other hand, SN remnants  hardly survive over time-scales longer than $10^4$ yr in Milky Way like nuclei \citep[e.g][]{Rimoldi2015}. Given that very few SN-EMRIs are expected to complete their inspiral within such time period (see Fig.~\ref{fig:tinfall}) we suggest that it would be very unlikely to detect a SN-EMRI and to also identify the remnant of the SN that induced it.

\item If we assume that BHs receive substantially smaller natal kicks than NSs, we expect the vast majority of SN-EMRIs to involve NSs rather than BHs. In particular, in the assumption of slow kicks and a TE distribution, the SN-EMRIs are NSs in the $\gtrsim 99\%$ of cases, and BHs only in the $\lesssim 1\%$ of cases (while the underlying CO mass functions counts roughly the same number of NSs and BHs). In contrast, if we assume a TE distribution and slow kicks, SN-EMRIs are NSs in the $60\%$ of cases and BHs in the remaining $40\%$. 

This aspect might help us to constrain the formation channel of EMRIs and to disentangle relaxation-driven EMRIs from SN-EMRIs. In fact, most relaxation-driven EMRIs are expected to be BHs. In contrast, SN-EMRIs involve NSs more likely than BHs.

\end{itemize}

So far, we solely presented the fraction of SNe that will result in a SN-EMRI. This number could be translated in a rate of SN-EMRIs per Milky Way galaxy per year. The easiest approach is to directly use the  rate at which SNe occur in the GC. Such rate has been suggested to be of the order of $10^{-4}$ yr$^{-1}$ both from theoretical arguments \citep{Zubovas2013,Rimoldi2015} and from the observation of a $\sim 10^4$ yr old SN remnant spotted within the SMBH sphere of influence \citep{Maeda2002}.




Consequently, the rate of SN-EMRIs per GC can be obtained by multiplying the EMRI fractions in Tab.~\ref{tab:emris} to $\sim 10^{-4}$ yr$^{-1}$, yielding values up to a few $\times 10^{-8}$ yr$^{-1}$ per Milky Way\footnote{ \citet{Babak2017} highlighted that SMBHs might outgrow their present-day masses if one assumes them to accrete continuously at the rates predicted by the most optimistic EMRI models \citep[e.g.][]{Amaro-Seoane2011}; here we can do the same exercise estimating how much a SMBH is expected to grow as a result of SN-driven EMRIs. Given that SN-induced plunges and EMRIs are expected to occur at a rate of a few $ \times 10^{-8}$ yr$^{-1}$ per Milky Way, and each accretion event would optimistically  increase the SMBH mass by 10 $\msun$,  we expect the SMBH to accumulate a few $\times 10^3 \msun$ in a Hubble time. Thus, reassuringly,  SN-induced EMRIs and plunges would not increase the mass of a Milky-Way sized SMBH by a considerable amount over the age of the Universe. }. 
 The rate of EMRIs per Milky Way induced by standard NRR processes is still not well constrained, but a typically adopted rate is a few$\times10^{-7}$ yr$^{-1}$ \citep[assuming a non-spinning SMBH and strong mass segregation, ][]{Amaro-Seoane2011}. Thus SN-EMRIs would constitute up to $10\%$ of traditional NRR induced EMRIs.

If we consider that the density of Milky-Way-like galaxies in the local Universe is $\sim{}0.0116$ Mpc$^{-3}$ \citep{Kopparapu2008,Abadie2010} and if we conservatively assume that only Milky-Way-like galaxies can produce SN-EMRIs,  a SN-EMRI rate of $\sim 10^{-8}$ yr$^{-1}$ per Milky Way-like galaxy translates into a SN-EMRI rate density $R_{\rm SN-EMRI}\sim{}0.116$ Gpc$^{-3}$ yr$^{-1}$.  
{ If we assume that the LISA mission will be able to detect EMRIs up to $z \approx{} 1$ (pessimistic assumption) or even up to $z \approx 3$ (optimistic assumption, \citealt{Babak2017}), we can expect it to observe  $\approx{} 4$ and  $\approx{} 30$ SN-EMRIs yr$^{-1}$ in the pessimistic and optimistic case, respectively. }
This prediction confirms that SN-EMRIs can play a major role in the landscape of GW sources, and deserves further investigation in preparation for the LISA mission.


\section*{Acknowledgements}

We warmly thank the anonymous referee for their very useful comments and suggestions. We  thank and all the members of the ForDys group for  stimulating discussion, especially M. Pasquato, A. Ballone and A. Trani; we also thank M. Arca-Sedda, R. Sahari, Z. Haiman, T. Tamfal and L. Marafatto for  constructive comments and suggestions. EB is pleased to thank the organizers and participants to the Astro-GR 2017 meeting, where enlightening talks and discussion sections first planted the seed that led to the present work.  %
We acknowledge financial support from the Istituto Nazionale di Astrofisica (INAF) through a Cycle 31st PhD grant, from the Italian Ministry of Education, University and Research (MIUR) through grant FIRB 2012 RBFR12PM1F and from INAF through grant PRIN-2014-14. EB acknowledges support  from the \textit{Fondazione Ing. Aldo Gini} and from the Swiss National Science Foundation under the Grant 200020\_178949. MM acknowledges financial support by the European Research Council for the ERC Consolidator grant DEMOBLACK, under contract no. 770017,  from the MERAC
Foundation through grant `The physics of gas and protoplanetary
discs in the Galactic centre', from INAF through PRIN-SKA `Opening a new era in pulsars and compact objects science with MeerKat', from MIUR through Progetto Premiale `FIGARO' (Fostering Italian Leadership in the Field of Gravitational Wave Astrophysics)
and `MITiC' (MIning The Cosmos Big Data and Innovative Italian
Technology for Frontier Astrophysics and Cosmology), and from
the Austrian National Science Foundation through FWF stand-alone
grant P31154-N27 `Unraveling merging neutron stars and black
hole-neutron star binaries with population-synthesis simulations'.




\bibliography{bibliography} 


\appendix

\section{Angular momentum NRR time}
\label{sec:appA}

\begin{figure}
\centering
\includegraphics[ trim={2mm 0 4mm 0},clip,width=0.37\textwidth]{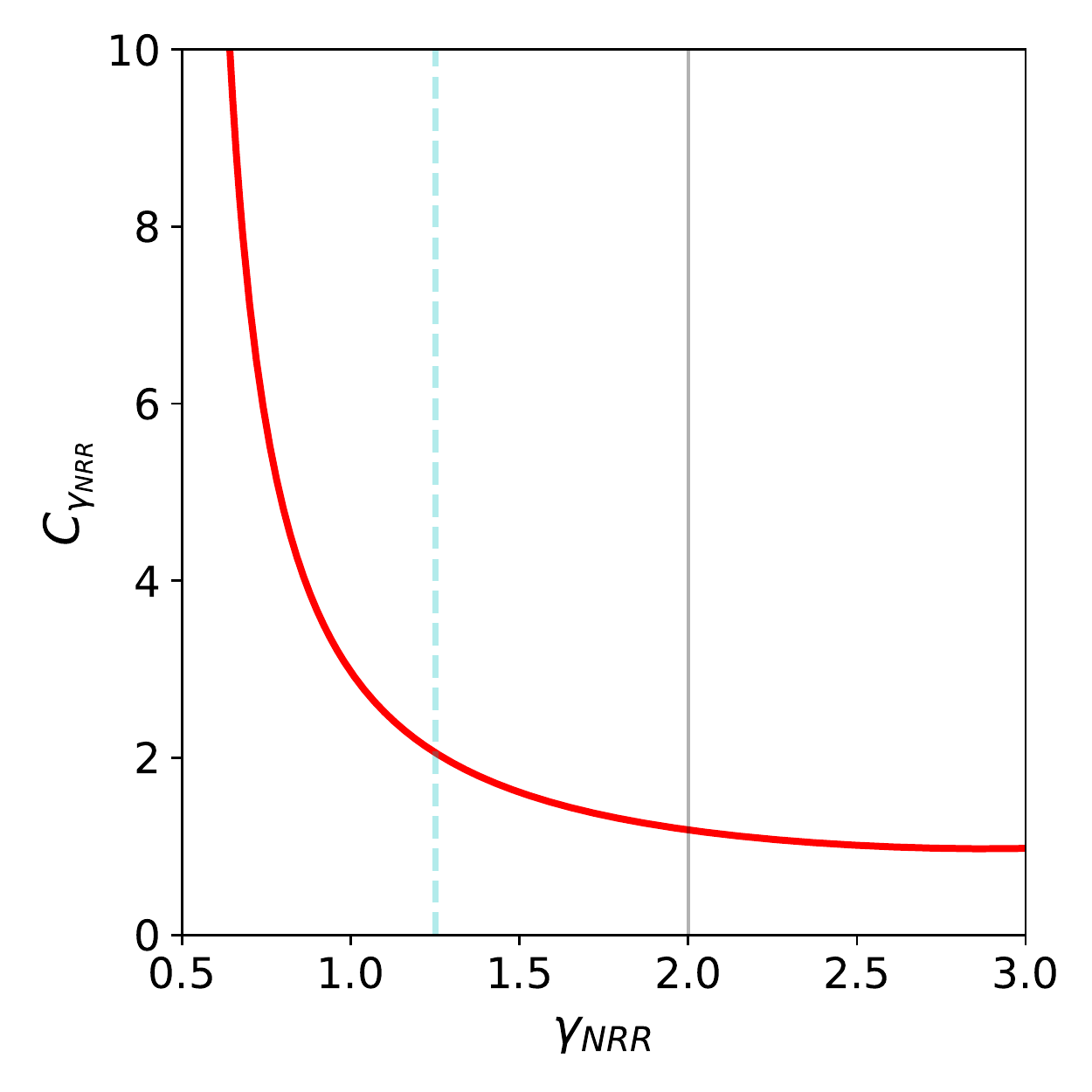} 
    \caption{%
   Value of $C_{\gamma_{\rm NRR}}$ (adopted for the evaluation of the angular-momentum NRR timescale in Eq.~\ref{eq:trel3}) as a function of $\gamma_{\rm NRR}$  (i.e. the slope of the assumed power-law density profile, $n_{\rm NRR} (r) \propto r^{-\gamma_{\rm NRR}}$).
The dashed cyan and solid grey lines mark the values of $\gamma_{\rm NRR}$ adopted in the present work for the background population (1.25 and 2); 
as a reference, $C_{\gamma_{\rm NRR}}$ = $\{ 5.74, 2.97, 2.06, 1.61, 1.35, 1.19, 1.01, 0.98\}$ 
for 
$\gamma_{\rm NRR}$ = $\{0.75, 1, 1.25, 1.5, 1.75, 2, 2.5, 3 \}$.}\label{fig:Cgamma}
\end{figure}

The derivation of the NRR timescale associated to angular momentum variations near a SMBH (Eq.~\ref{eq:trel3}) can be obtained following e.g. \citet{Cohn1978} and \citet{Milosavljevic2003}. In what follows we assume that stars distribute isotropically and homogeneously about the SMBH, and that the potential well is dominated by the SMBH presence. 
First of all, we define the relative energy as $\mathcal{E} = -v^2/2 + \psi(r) = -GM_\bullet/(2a)$, where $\psi(r)=GM_\bullet/r$; for consistency with the literature on the topic, we also set the angular momentum related variable $\mathcal{R} = j^2/j^2_c = 1 - e^2$. Given the new definitions, we can relate the number-density of stars in the $(\mathcal{E,R})$ plane to the phase-space distribution function $f$ as
\begin{equation}
N\mathcal{(E, R, }t)\  d\mathcal{E}\  d\mathcal{R} = 4\pi^2\  P(\mathcal{E}) \ j_c^2(\mathcal{E}) \ f\mathcal{(E, R, }t) \  d\mathcal{E} \  d\mathcal{R},
\end{equation}
where $P(\mathcal{E})$ is the  orbital period of a star with energy $\mathcal{E}$.  The Fokker-Plank equation that describes the diffusion in angular momentum can be written as
\begin{equation}
\frac{\partial N}{\partial t} = \frac{1}{2} \frac{\partial}{\partial \mathcal{R}} \left( \langle (\Delta \mathcal{R})^2 \rangle \frac{\partial N}{\partial \mathcal{R}} \right),
\end{equation}
where we used the relation \citep{Binney1988} 
\begin{equation}
\langle \Delta \mathcal{R} \rangle =  \frac{1}{2} \frac{\partial}{\partial \mathcal{R}} \langle (\Delta \mathcal{R})^2 \rangle
\end{equation}
between the first ($\langle \Delta \mathcal{R} \rangle$) and second order ($\langle (\Delta \mathcal{R})^2 \rangle$) Fokker Plank diffusion coefficients; $\langle (\Delta \mathcal{R})^2 \rangle$ represents the sum per unit time  of  $(\Delta \mathcal{R})^2 $ due to encounters. 

In the limit of $\mathcal{R}\longrightarrow 0$ and averaging over one orbital period, we can write
\begin{equation}
    \frac{\partial N}{\partial t} = \bar{w} \frac{\partial}{\partial \mathcal{R}} \left( \mathcal{R} \frac{\partial N}{\partial \mathcal{R}} \right)
\end{equation}
where 
\begin{equation}
\bar{w}(\mathcal{E}) = \frac{1}{P(\mathcal{E})} \oint \frac{dr}{v_r} \lim_{\mathcal{R} \to 0} \frac{\langle (\Delta \mathcal{R})^2 \rangle}{2 \mathcal{R}};
\end{equation}
in the previous equation, the integral is carried out over a radial period, $r$ represents the radial coordinate and $v_r$ the radial velocity. Following \citet{Hopman2005} and \citet{Merritt2011emris}, we define the NRR timescale as $t_r = \bar{w}^{-1}(\mathcal{E})$.

The phase-space density $f(\mathcal{E})$ can be related to $N(\mathcal{E})$ via
\begin{equation}\label{eq:f_of_E}
f(\mathcal{E}) = \frac{\mathcal{E}^{5/2} N(\mathcal{E}) }{\sqrt{2} \pi^3 (GM_\bullet)^3} = f_0 \mathcal{E}^{\gamma_{\rm NRR}-3/2}
\end{equation}
in the assumption that the number-density of stars in the physical space scales as $ a^{-\gamma_{\rm NRR}}$. We can express the local diffusion coefficient in terms of $f(\mathcal{E})$ via
\begin{subequations}
\label{eq:I_n}
\begin{eqnarray}
    \lim_{\mathcal{R} \to 0} \frac{\langle (\Delta \mathcal{R})^2 \rangle}{2 \mathcal{R}} = \frac{32 \ln \Lambda \left( \pi r G m_\star \right)^2 \left(3\mathscr{I}_{1/2} -\mathscr{I}_{3/2} +2\mathscr{I}_0 \right)}{3 j_c^2}, \\
    \mathscr{I}_0(\mathcal{E}) = \int_0^\mathcal{E} f(\mathcal{E}') d\mathcal{E}',\\
    \mathscr{I}_{n/2}(\mathcal{E},r) = \left[\psi(r) -\mathcal{E}\right]^{-n/2} \int_\mathcal{E}^\psi \left[\psi(r) -\mathcal{E}'\right]^{n/2} f(\mathcal{E}') d\mathcal{E}',
\end{eqnarray}
\end{subequations}
$\ln \Lambda$ being the Coulombian logarithm  defined in Sec.~\ref{sec:2br}. The orbit-averages of the previously defined $\mathscr{I}$s are
\begin{equation}\label{eq:I_mean}
  \bar{\mathscr{I}}(\mathcal{E}) = \frac{1}{\sqrt 2} \int_0^{GM_\bullet/\mathcal{E}} \frac{ \mathscr{I}(\mathcal{E},r)r^2 dr}{\sqrt{\psi - \mathcal{E}}}.
\end{equation}
By numerically integrating Equations~\ref{eq:I_n}-\ref{eq:I_mean} we can obtain the $\gamma_{\rm NRR}$-dependent constant
\begin{equation}
    C_{\gamma_{\rm NRR}} = \frac{3\mathscr{I}_{1/2} -\mathscr{I}_{3/2} +2\mathscr{I}_0 }{f_0 \mathcal{E}^{\gamma_{\rm NRR}-4}(GM_\bullet)^3};
\end{equation}
$C_{\gamma_{\rm NRR}}$ as a function of $\gamma_{\rm NRR}$ is plotted in Fig.~\ref{fig:Cgamma}. 
$C_{\gamma_{\rm NRR}}$ can be used to  express  $\bar{w}(\mathcal{E})$ as
\begin{equation}
    \bar{w}(\mathcal{E}) = \frac{64\pi \sqrt2 C_{\gamma_{\rm NRR}}}{3}{G^2m^2_\star \ln \Lambda f(\mathcal{E})}.
\end{equation}
The phase-space distribution $f(\mathcal{E})$ can be expressed as a function of the semimajor-axis $a$ as
\begin{equation}
f(a) = \frac{1}{4\pi^3}\frac{ N_0 a^{-\gamma_{\rm NRR}+3/2}}{ (GM_\bullet)^{3/2}};     
\end{equation}
 here $N_0$ is the normalizing constant to the number $N_{\rm NRR}$ of stars within a given $a$, i.e. $N_{\rm NRR}(<a)=N_0a^{3-\gamma_{\rm NRR}}$, as in Sec.~\ref{sec:2br}. Finally, Eq.~\ref{eq:f_of_E} coupled with 
 \begin{equation}
     N(\mathcal{E,R}) d\mathcal{E}d\mathcal{R}= N(a,e) dade
 \end{equation}
 allows us to express the NRR timescale as
 \begin{equation}
     t_r = \bar{w}^{-1}(a) =\frac{3\sqrt{2}\pi^2}{32 \,{}C_{\gamma_{\rm NRR}}} \left( \frac{G\,{}M_\bullet}{a}\right)^{3/2} \frac{a^{\gamma_{\rm NRR}}}{G^2 m^2_\star N_0 \ln \Lambda}.
 \end{equation}
This appendix is based on appendix B of \citet{Merritt2011emris}; a derivation similar to ours can also be found in \citet{Hamers2014}.


\bsp	
\label{lastpage}
\end{document}